\pdfoutput=1
\documentclass[cernpreprint, texlive=2016, UKenglish, texmf]{atlasdoc}
 
\usepackage{atlaspackage}
\usepackage{atlasbiblatex}
 
\usepackage{atlasphysics}
 
\graphicspath{{logos/}{figures/}}
 
\usepackage{ANA-TOPQ-2019-28-PAPER-defs}
\usepackage{multirow}
\usepackage{siunitx}
\usepackage{collcell}
 
\RequirePackage{atlassnippets}

\AtlasTitle{Measurement of the energy asymmetry in $t\bar{t}j$ production at $13\,$TeV  with the ATLAS experiment and interpretation in the SMEFT framework}
 
\AtlasAbstract{
A measurement of the energy asymmetry in jet-associated top-quark pair production is presented using $139\,\mathrm{fb}^{-1}$ of data collected by the ATLAS detector at the Large Hadron Collider during $pp$ collisions at $\sqrt{s}=13\,\text{TeV}$. The observable measures the different probability of top and antitop quarks to have the higher energy as a function of the jet scattering angle with respect to the beam axis. The energy asymmetry is measured in the semileptonic $t\bar{t}$ decay channel, and the hadronically decaying top quark must have transverse momentum above $350\,\text{GeV}$.
The results are corrected for detector effects to particle level in three bins of the scattering angle of the associated jet. The measurement agrees with the SM prediction at next-to-leading-order accuracy in quantum chromodynamics in all three bins. In the bin with the largest expected asymmetry, where the jet is emitted perpendicular to the beam, the energy asymmetry is measured to be $-0.043\pm0.020$, in agreement with the SM prediction of $-0.037\pm0.003$. Interpreting this result in the framework of the Standard Model effective field theory (SMEFT), it is shown that the energy asymmetry is sensitive to the top-quark chirality in four-quark operators and is therefore a valuable new observable in global SMEFT fits.
}
 
\author{The ATLAS Collaboration}
 
\AtlasRefCode{TOPQ-2019-28}
 
\AtlasJournalRef{Eur. Phys. J. C 82 (2022) 374}
\AtlasDOI{10.1140/epjc/s10052-022-10101-w}
 
\PreprintIdNumber{CERN-EP-2021-181}

\hypersetup{pdftitle={ATLAS document},pdfauthor={The ATLAS Collaboration}}
 
\begin{document}
 
\maketitle
 
\tableofcontents

\section{Introduction} 
In searches for physics beyond the Standard Model (SM) at the Large Hadron Collider (LHC), quantum effects of virtual new heavy particles in high-energy observables play a crucial role. In the framework of the Standard Model effective field theory (SMEFT), such effects can be described by effective couplings that modify and extend the interactions of SM particles at energies within the reach of the LHC~\cite{Buchmuller:1985jz,Leung:1984ni,Brivio:2017vri}. Extensions of the Standard Model at high energies, referred to as UV (`ultraviolet') theories, typically leave a pattern of effects in several observables. To pin down the structure of a UV theory, sensitive observables that probe complementary directions in the large parameter space of effective couplings have to be combined in a global analysis.
 
Top-quark physics at the LHC is very well suited to deciphering possible UV theories, both in sensitivity and in complementarity. Top-quark pair production provides a large number of precise observables, allowing the detection of even small deviations from SM predictions in the presence of new physics. Among these observables, the charge asymmetry,
based on the different probability for top and antitop quarks to be emitted in a considered phase-space region, is particularly sensitive to new physics. In QCD, the charge asymmetry in $t\bar t$ production first occurs at next-to-leading order (NLO)~\cite{Kuhn:1998kw}; in $t\bar t j$ production it is induced at tree level due to the presence of the additional jet~\cite{Bowen:2005ap,Dittmaier:2007wz,Dittmaier:2008uj,Melnikov:2010iu}. Contributions from new particles can lead to significant modifications of the asymmetry compared to the SM prediction~\cite{Antunano:2007da,Bauer:2010iq,Haisch:2011up,Kamenik:2011wt}.
 
At the LHC, the charge asymmetry has been measured as a rapidity asymmetry\footnote{In the literature, `charge asymmetry' is often used as a synonym for `rapidity asymmetry'. This paper distinguishes between `charge asymmetry' as a feature of $t\bar t$ production in a theory, and `rapidity asymmetry' and `energy asymmetry' as observables sensitive to this feature.} in top-quark pair production and is found to be in good agreement with the SM prediction~\cite{TOPQ-2016-16,TOPQ-2014-16,CMS-TOP-12-033}.
Alternatively, the charge asymmetry can be measured as an \textit{energy asymmetry} in top-quark pair production in association with a high transverse momentum (high $\pt$) jet~\cite{Berge:2013xsa}. The energy asymmetry is an important test of the charge asymmetry at leading order in QCD. By construction, the energy asymmetry is sensitive to the charge asymmetry in a different phase-space region than the rapidity asymmetry. In combination, the two asymmetries are therefore powerful probes of physics beyond the Standard Model in SMEFT.
In this article, the energy asymmetry is measured with the $13\tev$ proton-proton collision data collected by the ATLAS experiment during Run 2 in 2015-2018 of the LHC.
 
The energy asymmetry in $\ttj$ production is defined as
\begin{equation}
\label{eq:ea-opt:explicit}
A_E(\thetaj) \equiv \frac{ \sigma^\text{opt}(\thetaj | \DeltaE > 0) - \sigma^\text{opt}(\thetaj | \DeltaE < 0)}{ \sigma^\text{opt}(\thetaj | \DeltaE > 0) + \sigma^\text{opt}(\thetaj | \DeltaE < 0) }\,,
\end{equation}
with the \emph{optimised cross section}
\begin{align} \label{eq:sigma-opt}
\sigma^\text{opt}(\thetaj) =  \sigma(\thetaj | y_{t\bar{t}j} > 0) + \sigma(\pi - \thetaj | y_{\ttj} < 0)\,,\qquad \theta_j \in [0,\pi]\,,
\end{align}
where $\sigma(\theta_j)$ is the differential cross section as a function of $\theta_j$, and $y_{\ttj}$ is the rapidity of the \ttj system.
 
\begin{figure}
\centering
\subfloat[]{\includegraphics[width=.40\textwidth]{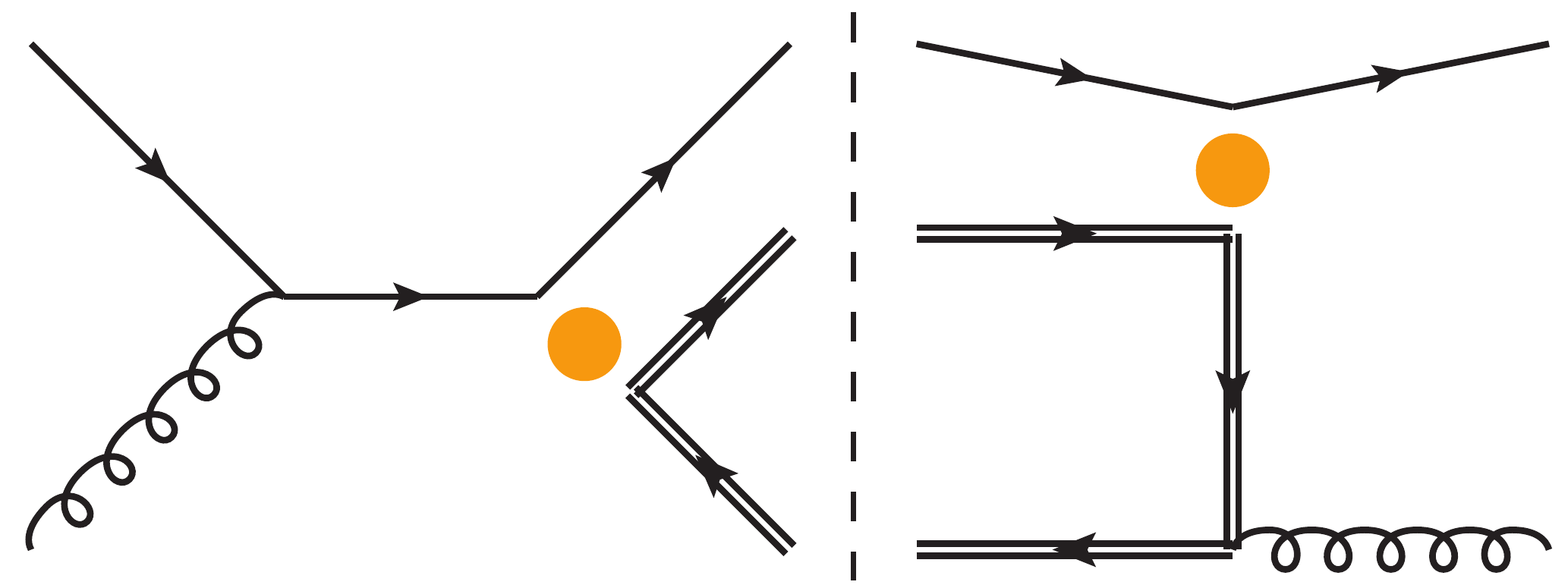}} \hspace*{1cm}
\subfloat[]{\includegraphics[width=.43\textwidth]{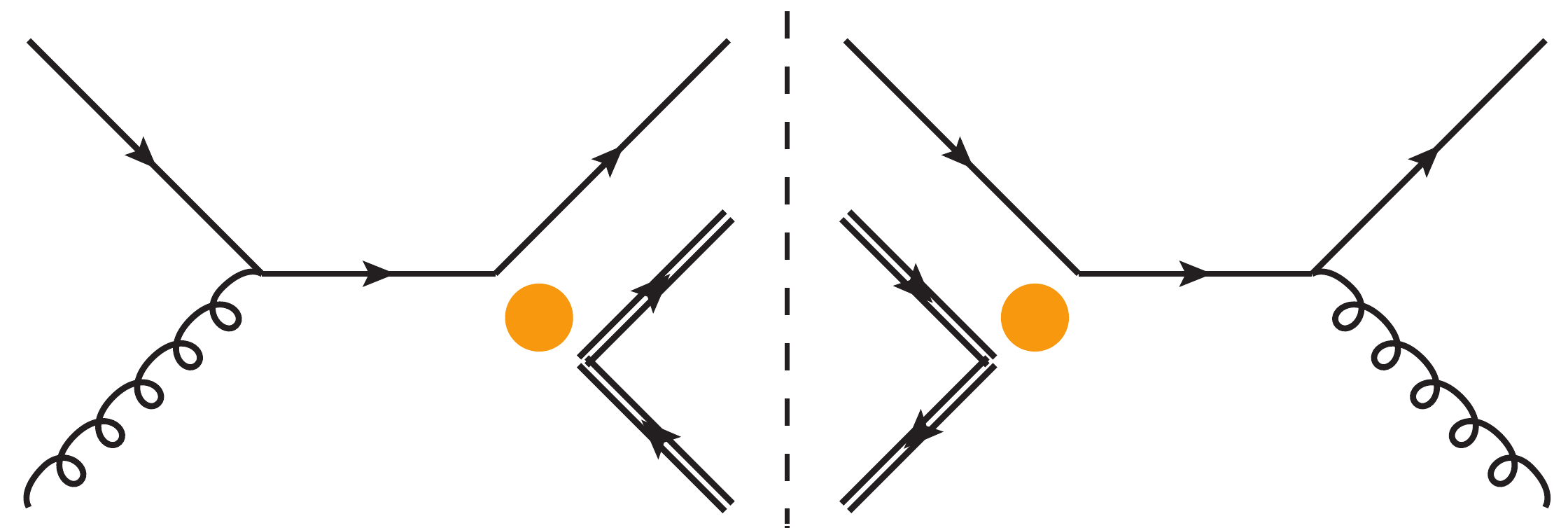}}
\caption{Examples of contributions to the energy asymmetry in terms of Feynman diagrams. The double lines indicate top quarks; the orange filled circles stand for possible insertions of four-quark operators in SMEFT. The dashed line symbolizes the interference $\mathcal{M}_1 \mathcal{M}_2^*$ of the two $qg \to \ttbar{q}$ amplitudes $\mathcal{M}_1$ and $\mathcal{M}_2^*$ to its left and right sides. Contributions to the asymmetry in QCD are obtained by replacing the orange filled circles in the left diagram by gluons. Further details about the QCD prediction and SMEFT contributions can be found in Ref.~\cite{Berge:2016oji} and Ref.~\cite{Basan:2020btr}, respectively.}
\label{fig:feynman}
\end{figure}

To measure the energy asymmetry the difference of the top and antitop energies $\DeltaE = E_t-E_{\bar t}$ is determined as a function of the jet angle $\thetaj$. Both $\DeltaE$ and $\thetaj$ are defined in the \ttj rest frame, which corresponds to the partonic centre-of-mass frame in tree-level processes. The angle $\thetaj$ is defined as the angle between the jet direction and the positive $z$-axis, i.e., the direction of parton $p_1$ in the process $p_1 p_2 \to \ttj$. The orientation of the $z$-axis can be chosen in either direction along the beam line, due to the symmetry in proton-proton collisions. The energy asymmetry is mainly generated in the partonic process $q g \to t\bar{t} q$, as illustrated in Figure~\ref{fig:feynman}. The outgoing quark-jet is boosted in the direction of the incoming valence quark. This boost is reflected in the rapidity of the \ttj system in the laboratory frame, $y_{t\bar t j}$. By combining `forward' events having $y_{\ttj} > 0$ with `backward' events having $y_{\ttj} < 0$ in the optimised cross section $\sigma^\text{opt}(\theta_j)$, the statistical sensitivity to the energy asymmetry is optimised~\cite{Alte:2014toa,Berge:2016oji}.
 
The energy asymmetry in $\ttj$ production is measured in three bins of the jet angle \thetaj in a fiducial phase space defined at particle level. The analysis selects events with a high-$\pt$ jet, one leptonic $W$ decay from one of the top quarks (called in the following the leptonically decaying top quark) and one hadronic $W$ decay from the other top quark (called in the following the hadronically decaying top quark). The decay products of the hadronically decaying top are required to be collimated in one large-radius jet, as is characteristic of the boosted regime. By focusing on this boosted regime, the additional jet is easily distinguished from the top-quark decay products. Moreover, the energy asymmetry increases with the transverse momentum of the associated jet~\cite{Berge:2016oji}.
 
The energy asymmetry is highly sensitive to the chirality of the top and antitop quarks~\cite{Basan:2020btr}, which allows the detection of even small potential deviations from the SM prediction. In SMEFT, such effects are described by four-quark operators that involve quarks with different chiralities. The detection of the additional jet in $t\bar t j$ production modifies the relative contributions of effective operators to the observables. As a consequence, observables in $\ttj$ production probe the parameter space of effective couplings in new directions beyond those sensitive to the rapidity asymmetry and other $\ttbar$ observables~\cite{Basan:2020btr}. The energy asymmetry is therefore a valuable new observable in global SMEFT fits~\cite{Buckley:2015lku,Hartland:2019bjb,Brivio:2019ius,Ellis:2020unq,Ethier:2021bye}, and thus the results of this measurement are interpreted in SMEFT to derive bounds on four-quark operator contributions.
 
This paper is organised as follows: a short description of the ATLAS detector is given in Section~\ref{sec:detector}, followed by the object definitions at detector and particle level in Section~\ref{sec:objects}; event selection and reconstruction are discussed in Section~\ref{sec:event}, while the simulated samples used to model signal and background events are described in Section~\ref{sec:samples}; Sections \ref{sec:unfolding} and \ref{sec:syst} deal with the unfolding method and the considered systematic uncertainties, respectively; the measurement results are presented in Section~\ref{sec:SMresult} and interpreted in the context of SMEFT in Section~\ref{sec:smeft}. Section~\ref{sec:summary} provides a summary of the analysis.


\section{ATLAS detector} 
\label{sec:detector}
The ATLAS detector~\cite{PERF-2007-01} at the LHC covers nearly the entire solid angle around the collision point.\footnote{\AtlasCoordFootnote}
It consists of an inner tracking detector surrounded by a thin superconducting solenoid, electromagnetic and hadron calorimeters,
and a muon spectrometer incorporating three large superconducting air-core toroidal magnets.
 
The inner-detector system (ID) is immersed in a \SI{2}{\tesla} axial magnetic field
and provides charged-particle tracking in the range \(|\eta| < 2.5\).
The high-granularity silicon pixel detector covers the vertex region and typically provides four measurements per track,
the first hit normally being in the insertable B-layer installed before Run~2~\cite{ATLAS-TDR-19,PIX-2018-001}.
It is followed by the silicon microstrip tracker, which usually provides eight measurements per track. These silicon detectors are complemented by the transition radiation tracker,
which enables radially extended track reconstruction up to \(|\eta| = 2.0\).
It also provides electron identification information
based on the fraction of hits (typically 30 in total) above a higher energy-deposit threshold corresponding to transition radiation.
 
The calorimeter system covers the pseudorapidity range \(|\eta| < 4.9\).
Within the region \(|\eta|< 3.2\), electromagnetic calorimetry is provided by barrel and
endcap high-granularity lead/liquid-argon (LAr) calorimeters,
with an additional thin LAr presampler covering \(|\eta| < 1.8\)
to correct for energy loss in material upstream of the calorimeters.
Hadron calorimetry is provided by the steel/scintillator-tile calorimeter,
segmented into three barrel structures within \(|\eta| < 1.7\), and two copper/LAr hadron endcap calorimeters.
The solid angle coverage is completed with forward copper/LAr and tungsten/LAr calorimeter modules
optimised for electromagnetic and hadronic energy measurements respectively.
 
The muon spectrometer (MS) comprises separate trigger and
high-precision tracking chambers measuring the deflection of muons in a magnetic field generated by the superconducting air-core toroidal magnets.
The field integral of the toroids ranges between \num{2.0} and \SI{6.0}{\tesla\metre}
across most of the detector.
A set of precision chambers covers the region \(|\eta| < 2.7\) with three layers of monitored drift tubes,
complemented by cathode-strip chambers in the forward region, where the background is highest.
The muon trigger system covers the range \(|\eta| < 2.4\) with resistive-plate chambers in the barrel, and thin-gap chambers in the endcap regions.
 
Interesting events are selected by the first-level trigger system implemented in custom hardware,
followed by selections made by algorithms implemented in software in the high-level trigger~\cite{TRIG-2016-01}.
The first-level trigger accepts events from the \SI{40}{\MHz} bunch crossings at a rate below \SI{100}{\kHz},
which the high-level trigger reduces in order to record events to disk at about \SI{1}{\kHz}.
 
An extensive software suite~\cite{ATL-SOFT-PUB-2020-001} is used in the reconstruction and analysis of real
and simulated data, in detector operations, and in the trigger and data acquisition systems of the experiment.

 
\section{Object definition} 
\label{sec:objects}
This section defines the final state objects at detector and particle levels. The particle-level objects are used to define the fiducial phase space for the measurement.
 
\subsection{Detector-level objects}
\label{sec:objects_detector}
The primary vertex of an event is the vertex which has the highest sum of squared transverse momenta from associated tracks with $\pt>0.5\gev$ \cite{ATL-PHYS-PUB-2015-026}. The primary vertex is required to have at least two tracks.
 
Electron candidates are reconstructed from clusters of energy deposits in the electromagnetic calorimeter that are matched to a track in the ID \cite{EGAM-2018-01}. They are required to have a transverse energy greater than $25\gev$ and $|\eta|<2.47$. Electron candidates are excluded if their calorimeter energy clusters lie within the transition region between the barrel and the endcap of the electromagnetic calorimeter ($1.37<|\eta|<1.52$). Electron tracks must satisfy $|z_0 \sin{\theta}|<0.5\,\mathrm{mm}$ and $|d_0|/\sigma(d_0)<5$, where $z_0$ is the longitudinal impact parameter relative to the reconstructed primary vertex, $d_0$ is the transverse impact parameter relative to the beam axis, and $\sigma(d_0)$ is the uncertainty in $d_0$. `Baseline' electron candidates are required to satisfy the `Loose' likelihood-based quality criterion of a multivariate algorithm~\cite{PERF-2017-01} based on shower shape and track selection requirements. `Tight' electron candidates are required to satisfy the `Tight' quality criterion and to pass $\pt$- and $\eta$-dependent isolation requirements based on their tracks and energy clusters, defined for the `Tight' isolation working point~\cite{PERF-2017-01}.
 
Muons are reconstructed by combining track information from the MS with tracks found in the ID. `Baseline' muon candidates must satisfy $\pt>25\gev$, $|\eta|<2.5$ and the `Medium' identification criterion defined in \Bibref{MUON-2018-03}. This criterion uses requirements on the number of hits in the various ID and MS subsystems and the $q/p$ compatibility between the MS and ID measurements. In addition, the track associated with the muon candidate must have $|z_0 \sin{\theta}|<0.5\,\mathrm{mm}$ and $|d_0|/\sigma(d_0)<3$. `Tight' muon candidates are furthermore required to be isolated from additional activity in the event according to the `Tight' isolation definition using particle-flow objects~\cite{MUON-2018-03}. 
 
Small-radius (\smallR) jets are reconstructed from particle-flow objects \cite{PERF-2015-09}, using the \akt jet clustering algorithm \cite{Cacciari:2008gp} as implemented in the FastJet package \cite{Fastjet}, with a radius parameter of $R=0.4$.
The impact of additional \pp\ interactions (pile-up) is mitigated by using the jet-area method to correct the jet \pt \cite{Cacciari:2008gn}.
Jets are calibrated to ensure a consistent jet energy scale and resolution between data and simulation as described in Ref.~\cite{JETM-2018-05}.
Only jet candidates with $\pt>25\gev$ and $|\eta|<2.5$ are considered in this analysis. To mitigate the impact of pile-up, an additional selection criterion using a likelihood-based ‘jet-vertex-tagger’ (JVT) discriminant is applied to jets with $\pt<60\gev$ and $|\eta|<2.4$~\cite{PERF-2014-03}.
 
\SmallR jets containing $b$-hadrons ($b$-jets) are identified ($b$-tagged) using the deep neural network (DNN) based algorithm, DL1r \cite{FTAG-2018-01}. This neural network is based on distinctive features of $b$-hadrons in terms of the impact parameters of tracks and the displaced vertices reconstructed in the inner detector. The inputs of the DL1r network also include discriminating variables constructed by a recurrent neural network, which exploits the spatial and kinematic correlations between tracks originating from the same $b$-hadron. A selection requirement on the DNN discriminant output is used to identify $b$-jets. The chosen value provides an $85\%$ average efficiency for identifying $b$-jets in simulated \ttbar events and $c$-jet and light-flavour jet rejection factors of 2.6 and 29, respectively.

Large-radius (\largeR) jets are reconstructed with the \antikt algorithm from locally calibrated topological cell clusters \cite{PERF-2014-07} with radius parameter $R=1.0$, and then trimmed using parameters $R_\text{sub}=0.2$ and $f_\text{cut}=0.05$ \cite{Krohn:2009th}. The jet mass resolution is improved by combining the jet mass measurement in the calorimeter with the measurement of the charged component of the jet within the ID \cite{JETM-2018-02}. The \pt and mass calibrations of \largeR jets are based on $13\tev$ data and simulation \cite{JETM-2018-02}. Only \largeR jets with $|\eta|<2.0$ are selected. A \largeR jet is considered to be $b$-tagged if there is at least one $b$-tagged \smallR jet within $\Delta R = 1.0$ of its axis.
 
\LargeR jets containing top quarks are identified (top-tagged) with a DNN using jet substructure variables~\cite{JETM-2018-03} as input. The top-tagger has an efficiency of 80\% for \largeR jets that are matched to a particle-level jet, which is in turn matched to a `truth' top-quark in the MC generator's event record, that has jet mass above $140\gev$ and at least one ghost-associated~\cite{Cacciari:2008gn} $b$-hadron.  The rejection factor for jets not matched to a top quark ranges from $\approx 10$ to $\approx 50$ for \largeR jets with transverse momenta of $2000\gev$ and $350\gev$, respectively. To avoid correlations with additional analysis-specific selection criteria (see Section \ref{sec:event}), $b$-tagging information is not included as input to the DNN tagger.

Ambiguities can arise from the independent reconstruction of electron, muon and jet candidates in the detector. A sequential procedure (overlap removal) is applied to avoid double-counting of physics objects, using the same algorithm as in \Bibref{TOPQ-2018-15}.
 
The missing transverse momentum vector $\metvec$ (with magnitude \met) is calculated from the negative vector sum of the \pt of all reconstructed electrons, muons and \smallR jets after overlap removal, as well as the \pt of soft activity comprising tracks not associated with those reconstructed objects \cite{PERF-2016-07}.

\subsection{Particle-level objects}
\label{sec:objects_particle}
Particle-level objects are defined according to ATLAS common practices for `truth' particle observable definitions as outlined in \Bibref{ATL-PHYS-PUB-2015-013}. These are specified as close as possible to the definition used for experimentally reconstructed objects using only physical particles that enter the detector to minimise model-dependent corrections. Particle-level objects in simulated \ttbar events are defined using quasi-stable particles with a mean lifetime $\tau > 30\,\text{ps}$. They are selected after hadronisation but before the interaction of these particles with the detector components. Particles from pile-up effects are not included.
 
Electrons and muons must originate (either directly or from leptonic $\tau$ decays) from the $W$ bosons from top-quark decays, without requiring a direct match with the parent boson, excluding electrons and muons from hadron decays. The 4-momenta of the `truth' charged leptons are modified (`dressed') by adding the 4-momenta of all radiated photons within a cone of size $\Delta R=0.1$ around the lepton, excluding photons from hadron decays, to take into account final-state photon radiation. \SmallR and \largeR particle-level jets are reconstructed with the same jet algorithms as at detector level using all stable particles, but excluding the charged leptons and neutrinos not from hadron decays and also the photons used to dress the charged leptons. The $b$-hadrons with $\pt>5\gev$ are included in the clusters forming stable-particle jets with their energies set to a negligible positive value (ghost-association)~\cite{Cacciari:2008gn}. Particle-level \smallR jets containing one or more of these $b$-hadrons are considered to originate from a $b$-quark and are called $b$-flavoured at particle level. A particle-level \largeR jet is considered to be $b$-flavoured if there is at least one $b$-flavoured \smallR jet within $\Delta R = 1.0$ of its axis. A particle-level \largeR jet is considered to be top-labelled if it is $b$-flavoured and the jet mass satisfies is greater than $140\gev$. Charged leptons and \smallR jets must have $\pt>25\gev$ and $|\eta|<2.5$. The particle-level missing transverse momentum is defined as the vector sum of the transverse momenta of all neutrinos found in the simulation history of the event, excluding those originating from hadron decays.
 
Particle-level objects are subject to an overlap removal procedure similar to that applied for the detector-level objects as described in \Bibref{TOPQ-2018-15}.

 
\section{Event selection and reconstruction} 
\label{sec:event}
This analysis focuses on events with one leptonically and one hadronically decaying top quark in the boosted topology plus an additional hard jet. In this topology, almost all decay products of the hadronically decaying top quark are expected to be collimated into a single \largeR jet. The signature for the leptonically decaying top quark is an isolated charged lepton (electron or muon), missing transverse momentum and a $b$-tagged jet.
 
Data used in this analysis were collected by the ATLAS detector between 2015 and 2018 at $\sqrt{s}=13~\TeV$. Only events for which all detector subsystems were operational are considered. The data set corresponds to an integrated luminosity of \lumi.
 
Only events passing a single-electron or single-muon trigger~\cite{TRIG-2018-05,TRIG-2018-01} are considered. The lowest-threshold triggers use isolation requirements to reduce the trigger rate. These have $\pt$ thresholds of $20\gev$ for muons and $24\gev$ for electrons in 2015 data, and $26\gev$ for both lepton types in 2016, 2017 and 2018 data. They are complemented by additional triggers with a higher \pt threshold of $50\gev$ but no isolation requirements in order to increase the event acceptance.

The following event selection criteria were optimised to yield the highest expected sensitivity to the energy asymmetry for jet emission perpendicular to the beam axis. If not stated otherwise, the same criteria are applied for detector-level objects and particle-level objects:
 
\begin{itemize}
\item \textbf{Lepton candidate ($\ell$)} -- Exactly one tight lepton (muon or electron) with $\pt > 27\gev$ is required. This lepton is considered the candidate to be the one originated by the top quark via the $W$ decay and is therefore called the lepton candidate ($\ell$). Events containing additional baseline leptons with $\pt>25\gev$ are rejected. At particle level, exactly one particle-level lepton with $\pt > 27\gev$ is required, and events containing an additional lepton with $\pt>25\gev$ are rejected.
 
\item \textbf{$\met$ and $\Wboson$ boson transverse mass (\mwt)} -- To suppress events containing a fake or non-prompt lepton at detector level, it is required that $\met\ge20\gev$ and $\met + \mwt \ge 60\gev$.\footnote{$\mwt=\sqrt{\pt^\ell \met (1-\cos\Delta\phi)}$, where $\Delta\phi$ is the angle between the lepton candidate and \metvec\ in the plane transverse to the beam axis.} At particle level, no $\met$ or $\mwt$ cuts are applied.
 
\item \textbf{Hadronic top candidate ($j_h$)} -- At least one top-tagged \largeR jet fulfilling $\pt > 350\gev$, and which is separated from the lepton, $\Delta\phi(j_h,\ell) > 1.0$, is required. From the \largeR jets fulfilling these conditions, the hadronic top candidate is defined as the one with the highest \pt. At particle level, the top-tagging criterion is replaced by the top-labelling criterion defined in Section~\ref{sec:objects_particle}, while the kinematic requirements remain unchanged.
 
\item \textbf{Leptonic top $b$-jet candidate ($j_l$)} -- At least one \smallR jet close to the lepton candidate, $\Delta R(j_l,\ell) < 2.0$, and isolated from the hadronic top candidate, $\Delta R(j_l,j_h) > 1.5$, is required. From the \smallR jets fulfilling these conditions, the leptonic top $b$-jet candidate is defined as the highest-\pt identified $b$-jet. If no identified $b$-jet fulfils these conditions, the highest-\pt \smallR jet is taken.
 
\item \textbf{Associated jet candidate ($j_a$)} -- At least one \smallR jet with $\pt > 100\gev$ separated from the large jet, $\Delta R(j_a,j_h) > 1.5$, and the lepton, $\Delta R(j_a,\ell) > 0.4$, and different from the leptonic jet candidate is required. From the \smallR jets fulfilling these conditions, the associated jet candidate is defined as the one with the highest \pt. The $\pt$ cut was chosen such as to optimise the sensitivity to the energy asymmetry in the central $\thetaj$ bin, which includes events where the jet is emitted perpendicular to the beam axis and the asymmetry is largest. The optimisation is a trade-off between increased asymmetries for higher $\pt$ values \cite{Berge:2016oji} and increased statistical precision for smaller $\pt$ values.
 
\item \textbf{Identified $b$-jet} -- At least one \smallR jet is required to be identified as a $b$-jet. The $b$-jet must either be the leptonic jet candidate $j_l$ or be close to the hadronic top candidate $j_h$ within $\Delta R < 1.0$. If two or more \smallR jets are identified as $b$-jets, the leptonic jet candidate $j_l$ must be identified as a $b$-jet and the hadronic top candidate must be close to an identified $b$-jet.
 
\end{itemize}
 
The missing transverse momentum is used as the estimate of the transverse components of the neutrino momentum. The longitudinal component of the neutrino momentum is obtained by requiring that the invariant mass of the lepton--neutrino system equals the $W$ boson mass. If the resulting quadratic equation has more than one real solution, the solution with the smaller absolute longitudinal momentum is taken. If no real solution is found, the $\met$ vector is varied minimally in length and direction to obtain one real solution.
 
The $W$ boson candidate from the leptonic top quark is defined as the 4-momentum sum of the lepton candidate and the reconstructed neutrino. The leptonic top candidate is defined as the 4-momentum sum of the lepton candidate $\ell$, reconstructed neutrino, and leptonic jet candidate $j_l$. The charge of the leptonic top candidate is given by the lepton charge, and the opposite charge is assigned to the hadronic top candidate.
 
Signal events that pass the selection at both the particle level and the detector level are classified as fiducial signal, while events that pass the detector-level selection but not the particle-level selection are classified as non-fiducial signal.

\section{Signal and background modelling} 
\label{sec:samples}
All signal and background processes are modelled using Monte Carlo (MC) simulations, with the exception of backgrounds with fake or non-prompt leptons, which are estimated from data (see Section~\ref{sec:background:non-prompt}).
All simulated samples used \textsc{EvtGen\,1.6.0}~\cite{EvtGen} to model the decays of heavy hadrons, with the exception of the background samples generated with \textsc{Sherpa}~\cite{Bothmann:2019yzt}. Most of the MC samples were passed through a full simulation of the detector response with the \textsc{Geant4} toolkit~\cite{Agostinelli:2002hh,SOFT-2010-01}.
The samples used to estimate modelling systematic uncertainties are either obtained by reweighting the default full simulation samples, or produced using the fast simulation software \textsc{Atlfast}-II~\cite{ATL-PHYS-PUB-2010-013}, where sample comparisons are always performed using the same type of detector simulation. To model additional $pp$ interactions from the same or neighbouring bunch crossings, the hard scattering events were overlaid with a set of minimum-bias interactions generated using \textsc{Pythia\,8}~\cite{Sjostrand:2014zea} and the \textsc{MSTW2008lo}~\cite{Martin:2009iq} parton distribution function (PDF) set with a set of tuned parameters called the A3~\cite{ATL-PHYS-PUB-2016-017} tune. Finally, the simulated MC events are reconstructed using the same software as the data.
 
\subsection{Signal modelling}
\label{sec:samples:ttsig}
All \ttbar samples are normalised to the inclusive production cross section calculated by \textsc{Top++\,2.0}~\cite{Czakon:2011xx} at next-to-next-to-leading order (NNLO) in QCD including the resummation of next-to-next-to-leading-logarithm (NNLL) soft-gluon terms~\cite{Beneke:2011mq,Cacciari:2011hy,Baernreuther:2012ws,Czakon:2012zr,Czakon:2012pz,Czakon:2013goa}. This corresponds to $\sigma(\ttbar) = \mathrm{832\pm51~pb}$ at a top-quark mass of $m_t = 172.5\gev$, which is assumed for all \ttbar samples unless stated otherwise. The uncertainties in the cross section due to the choice of PDF and $\alphas$ are calculated using the PDF4LHC prescription~\cite{Botje:2011sn} with the \textsc{MSTW2008nnlo} 68\% CL~\cite{Martin:2009iq,Martin:2009bu}, \textsc{CT10nnlo}~\cite{Lai:2010vv,Gao:2013xoa} and NNPDF2.3 5f FFN~\cite{Ball:2012cx} PDF sets; these uncertainties are added in quadrature to the effect of the scale uncertainty.
 
The nominal \ttbar\ events were generated with the \powhegbox~\cite{Frixione:2007nw,Nason:2004rx,Frixione:2007vw,Alioli:2010xd}~v2 generator which provides matrix elements (MEs) at NLO in the strong coupling constant \alphas , with the \textsc{NNPDF3.0nlo}~\cite{Ball:2014uwa} PDF and the \hdamp\ parameter\footnote{The \hdamp\ parameter controls the transverse momentum \pt\ of the first additional emission beyond the leading order (LO) Feynman diagram in the parton shower (PS) and therefore regulates the high-\pt\ emission against which the \ttbar\ system recoils.} set to $1.5 m_t$~\cite{ATL-PHYS-PUB-2016-020}. The functional form of the renormalisation and factorisation scales (\muR and \muF) was set to the nominal scale of $\sqrt{m_{t}^2 + p_{\textrm T}^2}$. The events were interfaced with \pythia.230 for the parton shower (PS) and hadronisation, using the A14 tune~\cite{ATL-PHYS-PUB-2014-021} and the \textsc{NNPDF2.3lo} PDF set.
 
Various alternative $\ttbar$ samples are used to estimate the systematic uncertainties associated with the modelling of $\ttbar$ production. To simulate more (less) initial-state radiation (ISR), $\mur$ and $\muf$ were varied independently by a factor of $0.5$ ($2.0$) in the ME and the up (down) Var3c variations~\cite{ATL-PHYS-PUB-2014-021} of the A14 tune were used.
Samples with the renormalisation scale for QCD emission in final-state radiation (FSR) altered by a factors of $0.5$ and $2.0$ were used as well. To test the dependence of the measurement on the PS and hadronisation model, \powhegbox was interfaced with \herwigseven.04~\cite{Bahr:2008pv,Bellm:2015jjp} instead of \pythia.230, using the H7UE tune~\cite{Bellm:2015jjp} and the \textsc{MMHT2014lo} PDF set \cite{Harland-Lang:2014zoa}.
The uncertainty in the matching between the ME and PS is evaluated with a sample similar to the nominal \ttbar signal sample but where the \hdamp parameter is set to $3 m_t$, as well as a sample based on \mgamc (referred to as \mgamcshort in the following)~\cite{Alwall:2014hca} interfaced with \pythia. The calculation of the hard-scattering uses \mgamcshort\,2.6.0 with the \textsc{NNPDF3.0nlo} PDF set. Events were interfaced with \pythia.230, using the A14 tune and the \textsc{NNPDF2.3lo} PDF. The shower starting scale has the functional form $\mu_{\mathrm q} = H_{\mathrm T}/2$~\cite{ATL-PHYS-PUB-2017-007}, where $H_{\mathrm T}$ is defined as the scalar sum of the \pT of all outgoing partons. The choice for $\muR$ and $\muF$ is the same as used with \powhegbox. Two dedicated samples with the same settings as in the nominal \ttbar signal sample (\powhegbox+ \pythia) have $m_t$ set to either $172\gev$ or $173\gev$ to study the impact of mass uncertainties.
 
The \ttj events for the SM and SMEFT predictions used in Sections~\ref{sec:SMresult} and~\ref{sec:smeft} were generated with \mgamcshort interfaced with \pythia.
The calculation of the hard-scattering uses \mgamcshort\,2.7.2 with the \textsc{NNPDF3.0nlo} PDF set. Events were interfaced with \pythia.244 for PS and hadronisation, using the A14 tune and the \textsc{NNPDF2.3lo} PDF set. The renormalisation and factorisation scales were chosen as $\muR=\muF= m_t$.
The SM prediction is obtained from matrix elements at NLO in the strong coupling constant \alphas. The SM--EFT interference and squared EFT--EFT contributions were generated separately at leading order in \alphas using the SMEFT@NLO~\cite{Degrande:2020evl} package. Scale uncertainties in the SM and SMEFT predictions are obtained from the envelope of nine different scale variations by varying $\muR$ and $\muF$ independently by factors of 0.5 and 2.0.

\subsection{Background modelling}
There are two sources of background events: events with electrons and muons produced in $W$ or $Z/\gamma^{\star}$ boson decays (prompt leptons, Section~\ref{sec:background:prompt}) and events that have one lepton arising from another source (fake and non-prompt leptons, Section~\ref{sec:background:non-prompt}). In the latter case, electrons and muons originate from heavy-flavour hadron decays, photon conversions, or improper reconstruction of other physics objects such as hadronic jets.
 
\subsubsection{Backgrounds with prompt leptons}
\label{sec:background:prompt}
Background contributions with prompt leptons originate from a wide range of processes, which are estimated from simulation using the samples described below. The systematic uncertainties in the modelling of these processes by the simulation are discussed in Section~\ref{sec:syst}.
 
Single-top $Wt$ associated production was modelled using the \powhegbox~\cite{Re:2010bp,Nason:2004rx,Frixione:2007vw,Alioli:2010xd}~v2 generator to provide MEs at NLO in \alphas, using the five-flavour scheme with the \textsc{NNPDF3.0nlo} PDF set. The functional form of  \muR\ and \muF\ was set to the nominal scale of $\sqrt{m_{t}^2 + p_{\textrm T}^2}$. The diagram removal scheme~\cite{Frixione:2008yi} was employed to treat the interference with \ttbar\ production~\cite{ATL-PHYS-PUB-2016-020}. Dedicated samples produced with a diagram subtraction (DS) scheme~\cite{Frixione:2008yi} are used to evaluate the uncertainty due to the treatment of the overlap with \ttbar\ production.
 
Single-top s-channel (t-channel) production was modelled using the \powhegbox~\cite{Frederix:2012dh,Nason:2004rx,Frixione:2007vw,Alioli:2010xd, Alioli:2009je}~v2 generator to provide MEs at NLO in \alphas\ , using the five-flavour (four-flavour) scheme with the \textsc{NNPDF3.0nlo} PDF set. The functional form of  \muR\ and \muR\ was set to $\sqrt{m_{b}^2 + p_{{\textrm T},b}^2}$~\cite{Frederix:2012dh}. For these processes, the events were interfaced with \pythia.230 using the A14 tune and the \textsc{NNPDF2.3lo} PDF set.
 
Single-top uncertainties due to ISR are estimated using varied weights in the ME and in the PS. To evaluate this \muR\ and \muF\ were varied by a factor of 2.0 up and down.
The impact of increased or decreased FSR is evaluated using PS weights, which vary the renormalisation scale for QCD emission in the FSR by a factor of 0.5 and 2.0, respectively.
The impact of the PS and hadronisation model is evaluated by comparing the nominal generator sample with events produced with the \powhegbox~v2 generator at NLO in QCD.  These use the five-flavour (four-flavour) scheme for $tW$ and s-channel (t-channel) processes, and the \textsc{NNPDF3.0nlo} PDF set.
The events are interfaced with \herwigseven.04, using the H7UE tune and the \textsc{MMHT2014lo} PDF set \cite{Harland-Lang:2014zoa}.
To assess the uncertainty due to the choice of matching scheme, the nominal sample is compared with a sample generated with the \mgamcshort\,2.6.2 generator at NLO in QCD in the five-flavour (four-flavour) scheme for $tW$ and s-channel (t-channel) processes, using the \textsc{NNPDF3.0nlo} PDF set.
The events were interfaced with \pythia.230, using the A14 tune and the \textsc{NNPDF2.3lo} PDF.

QCD $V$+jets (\Wjets and \Zjets) production was simulated with the \sherpa~v2.2.1 MC generator.
In this set-up, NLO MEs for up to two jets, and leading order (LO) MEs for up to four jets, were calculated with the OpenLoops~\cite{Cascioli:2011va,Denner:2016kdg}  and Comix~\cite{Gleisberg:2008fv} libraries. All \sherpa samples use the nominal \sherpa PS~\cite{Schumann:2007mg}, based on Catani--Seymour dipoles and the cluster hadronisation model~\cite{Winter:2003tt}, which employs a dedicated set of tuned parameters developed by the \sherpa for use with the \textsc{NNPDF3.0nnlo} PDF set.
The $V$+jets samples are normalised to a NNLO prediction~\cite{Anastasiou:2003ds}.
 
Diboson (\VV) samples were simulated with the \sherpa\,2.2.1 and \sherpa\,2.2.2 PS MC generators.
\sherpa\,2.2.2 was used for two- and three-lepton samples.
Matrix element and PS matching~\cite{Hoeche:2011fd} was applied to the various jet multiplicities, which were then merged into an inclusive sample using an improved CKKW matching
procedure~\cite{Catani:2001cc,Hoeche:2009rj}.  The procedure was extended to NLO using the MEPS@NLO prescription~\cite{Hoeche:2012yf}.
These simulations are at NLO for up to one additional parton and at LO for up to three additional parton emissions using factorised on-shell decays.
The virtual QCD corrections for matrix elements at NLO were provided by the OpenLoops library~\cite{Cascioli:2011va,Denner:2016kdg}.
The calculation was performed in the $G_\mu$ scheme, ensuring an optimal description of pure electroweak interactions at the electroweak scale.
 
The production of \ttV\ and \ttH\ events was modelled using the \mgamcshort\,2.3.3 and  \powhegbox generators, respectively. The generators provide matrix elements at NLO in \alphas, with the \textsc{NNPDF3.0nlo} PDF set. For \ttV\ and \ttH\ production, the events were interfaced with \pythia.210 and \pythia.230, respectively. Each uses the A14 tune and the \textsc{NNPDF2.3lo} PDF set.
 
\subsubsection{Backgrounds with fake or non-prompt leptons}
\label{sec:background:non-prompt}
To estimate the total contribution of events with fake or non-prompt leptons a data-driven ``matrix-method''~\cite{ATLAS-CONF-2014-058} is used.
Two categories of events are selected, satisfying `baseline' (identification only) and `tight' (identification and isolation) lepton selection requirements as described in Section~\ref{sec:objects_detector}. The real (fake/non-prompt) lepton efficiency, $\varepsilon_{\text{real}}$ ($\varepsilon_{\text{fake}}$), is defined as the ratio of the number of real (fake/non-prompt) leptons satisfying the tight selection to the number of real (fake/non-prompt) leptons satisfying the baseline selection.
 
The efficiencies for real and fake/non-prompt leptons are parametrised with respect to the $\pT$ and $|\eta|$ of the leptons and are measured separately for electrons and muons in control regions that are statistically independent of the signal region.
The fake/non-prompt efficiencies are estimated using single-lepton events from data, where criteria are imposed to suppress the contribution from real leptons. The requirement on the missing transverse momentum is reversed relative to the signal region, to $\met < 20 \gev$, to reduce the contribution from $W$ boson and \ttbar events. The residual contribution from real prompt-lepton events in the selected sample is estimated using a MC simulation and subtracted separately from both the tight and baseline samples before the ratio of these samples is calculated to measure the efficiency. The efficiency for fake/non-prompt electrons (muons) ranges from $23\%$ ($13\%$) to $59\%$ ($23\%$), depending on the lepton $\pT$, $|\eta|$ and data taking period. Any differences between the fake/non-prompt lepton efficiencies in the control region with $\met < 20\gev$ and a validation region with $\met > 20\gev$ have been found to be well within the normalisation uncertainty applied on fake/non-prompt lepton backgrounds as described in Section \ref{sec:syst}.
The prompt- or real-lepton efficiencies are evaluated from the simulated \ttbar sample, corrected with scale factors measured in data~\cite{PERF-2017-01,MUON-2018-03}, in a control region with zero $b$-tagged jets. The efficiency for real electrons (muons) ranges from $83\%$ ($77\%$) to $95\%$ ($95\%$), depending on the lepton $\pT$, $|\eta|$ and data taking period.
 
The sample of events with fake or non-prompt leptons is estimated from the weighted data events, where the weight depends on the real- and fake-lepton efficiencies. As the dominant source of these events is expected to be QCD multijet production, this background is hereafter referred to as `Multijet'.
 
\subsection{Event yields}
Table \ref{tab:selection:EventYield} summarizes the event yields for the signal and background predictions before unfolding (pre-marginalisation) in comparison with those observed in data. The expected yield is in agreement with the observed number of events within uncertainties. The \ttbar signal purity is $87\%$, out of which 70\% are fiducial signal. The non-fiducial signal consists mainly of events in which the hadronic top candidate does not pass the $\pt$, mass and $b$-flavour requirements.
The most important background is $W$ boson production in association with jets (\Wjets) which contributes at the level of 5\% due to the similarity of its final state and kinematics to the signal. Single-top production contributes at the level of 3\% and is dominated by the $Wt$-channel. The Multijet yield  fraction in the signal region is found to be below 2\%.
The uncertainty of 21\% in the \ttbar\ signal prediction originates mainly from signal modelling effects, where the corresponding impact on the signal yield cancels out for the energy asymmetry observables.
Most backgrounds are expected to be symmetric and this is seen within the statistical uncertainty of the simulated samples. For \Wjets events the expected asymmetry is negative due to the combination of a higher cross section for $W^+$ than for $W^-$ production and the selected phase space preferring hadronically decaying top candidates with a larger energy than the leptonically decaying ones, which mostly include the real $W$ bosons.
 
\begin{table}
\centering
\caption{Event yields for the simulated signal and background predictions with uncertainties as described in Section~\ref{sec:syst}, compared with the event yield in the signal region as observed in data. The uncertainties do not add up in quadrature to the total uncertainty, due systematic uncertainties being correlated between event samples.}
\vspace{10pt}
\label{tab:selection:EventYield}
\begin{tabular}{lr@{${}\pm{}$}r}
\toprule
& \multicolumn{2}{c}{Events} \\
\midrule
\ttbar fiducial & $18\,000$ & $3700$ \\
\ttbar non-fiducial & $8000$ & $1700$ \\
\Wjets & $1600$ & $860$ \\
Single-top $Wt$ & $680$ & $340 $\\
Multijet & $580$ & $300$ \\
\ttV & $400$ & $200$ \\
\Zjets & $210$ & $110$ \\
\VV & $168$ & $85$ \\
\ttH & $147$ & $74$ \\
Single-top t-channel & $82$ & $40$\\
Single-top s-channel & $8$ & $3$\\
\midrule
Total & $30\,000$ & $5300$ \\
\midrule
Data & \multicolumn{1}{r@{\phantom{${}\pm{}$}}}{27\,265}\\
\bottomrule
\end{tabular}
\end{table}
 
\begin{figure}
\centering
\subfloat[]{\includegraphics[width=.33\textwidth]{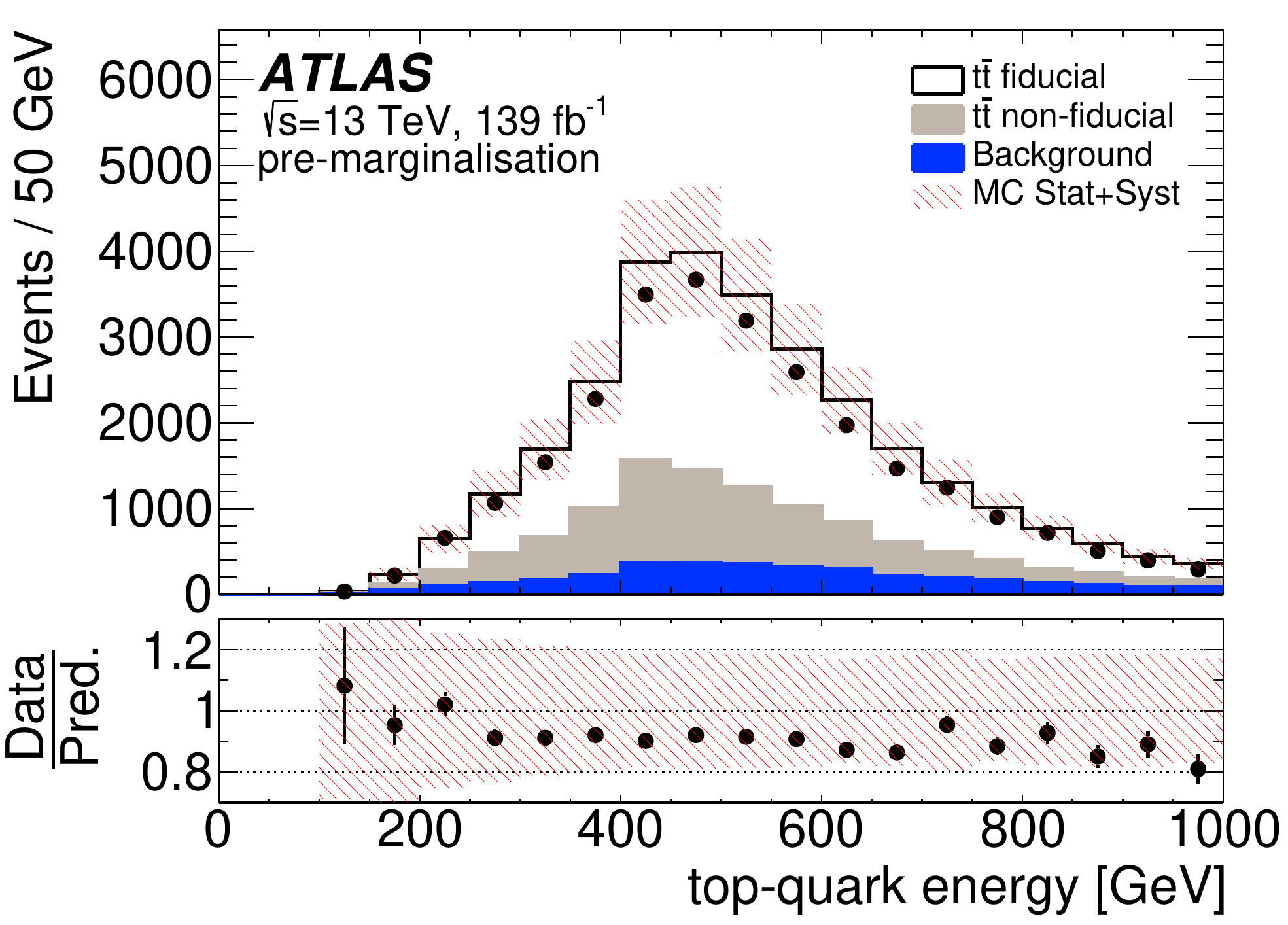}}
\subfloat[]{\includegraphics[width=.33\textwidth]{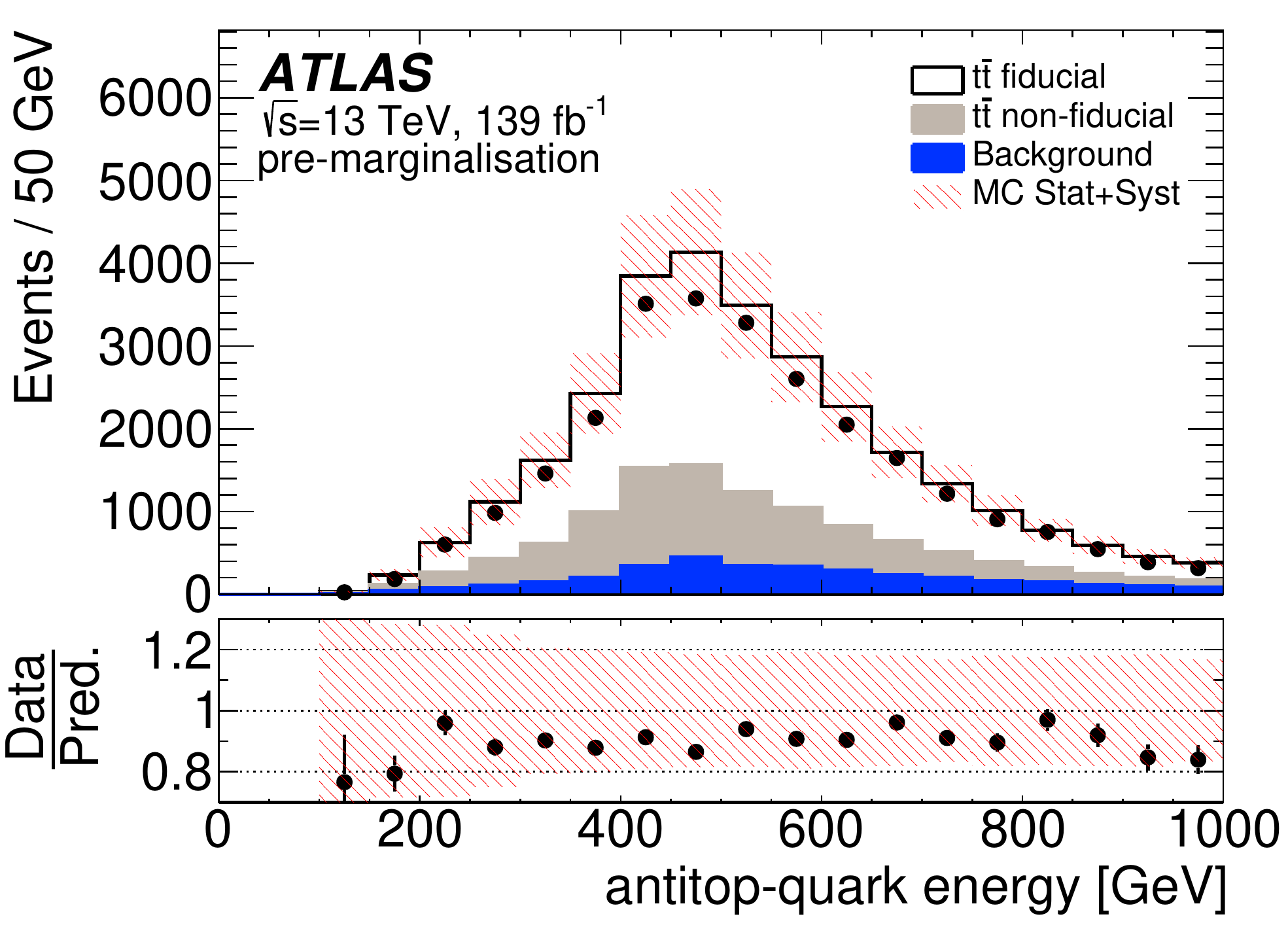}}
\subfloat[]{\includegraphics[width=.33\textwidth]{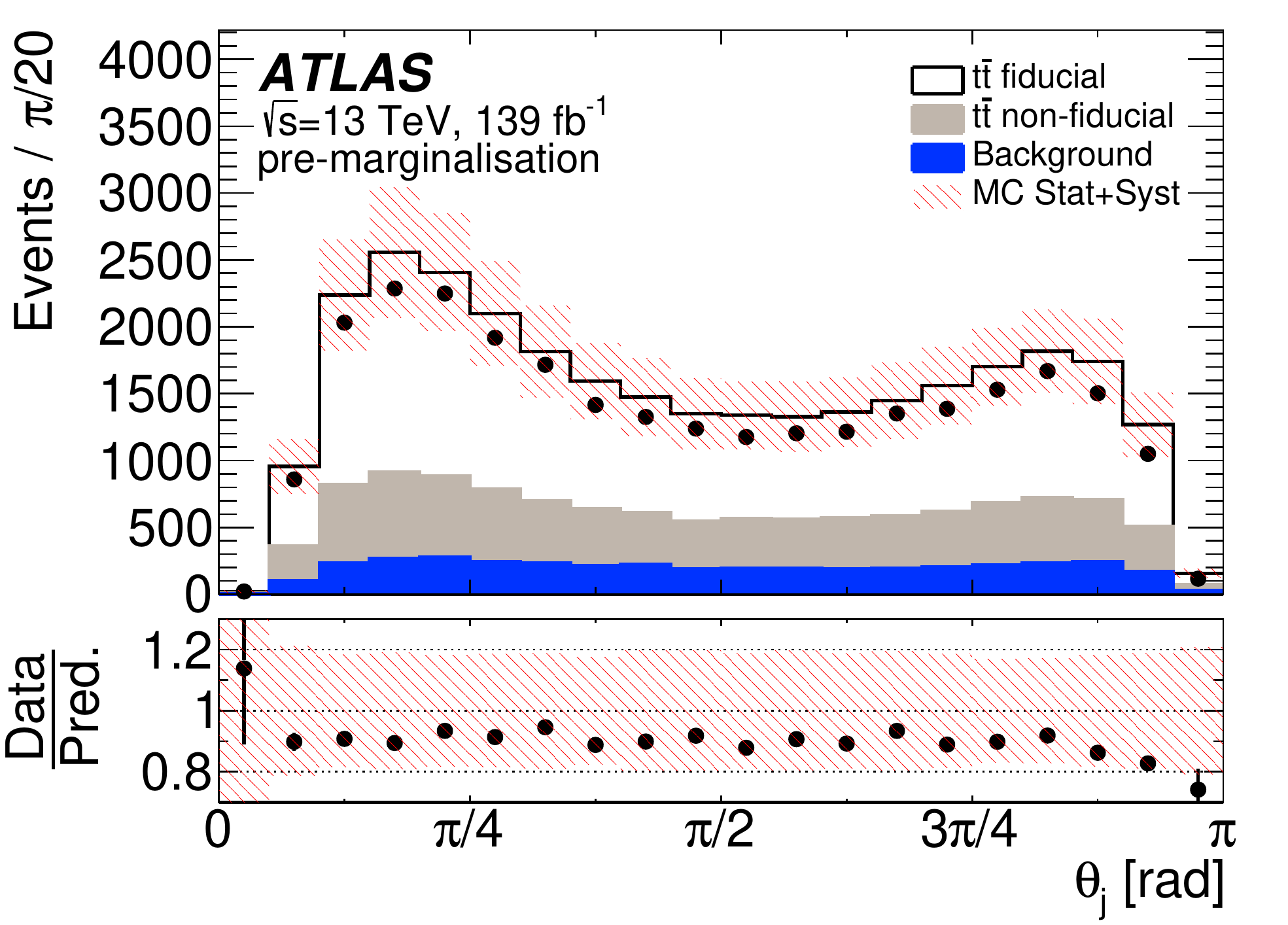}}
\caption{Comparison of observed data with the predicted distribution of (a) the top-quark energy, (b) the antitop-quark energy, and (c) the angle \thetaj before unfolding (pre-marginalisation).
The expectations for the $t\bar{t}$ fiducial and non-fiducial signal are represented by the white and grey areas, respectively, while all non-$t\bar{t}$ backgrounds are added up in the blue area. The shaded red bands correspond to the total uncertainty of the prediction, while the statistical uncertainty on the data is indicated by the black error bars, that are mostly smaller than the size of the marker.
}
\label{fig:data_prediction}
\end{figure}

Figure~\ref{fig:data_prediction} shows the level of agreement between the data and the detector-level prediction for the top-quark energy and the antitop-quark energy and \thetaj. All systematic uncertainties listed in Section~\ref{sec:syst} are included in these comparisons. The data are consistent with the predictions within uncertainties, with a predicted rate about 10\% larger than the observed data due to an overestimation of the $\ttbar$ yield in the boosted phase space, as reported in Refs.~\cite{TOPQ-2018-15,ATL-PHYS-PUB-2020-023}. As the charge asymmetry observable is independent of the $\ttbar$ cross-section, it is not sensitive to any mismodelling in the rate of $\ttbar$ production.

\section{Asymmetry extraction and unfolding} 
\label{sec:unfolding}
\begin{figure}[t!]
\centering
\includegraphics{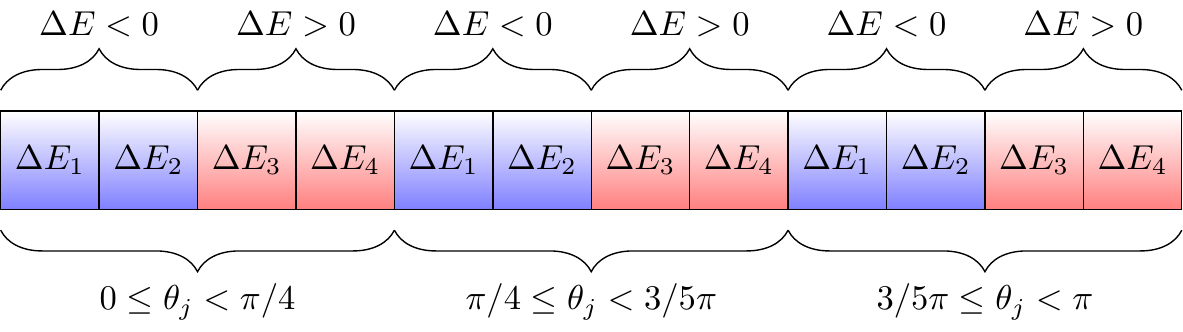}
\caption{Concatenation of optimised event rates $N^{\text{opt}}(\theta_j,\DeltaE)$ in three bins of $\theta_j$, $[0,\pi/4,3\pi/5,\pi]$, and four bins in $\DeltaE$, $[-\infty,-100,0,100,\infty]\gev$. For each bin in \thetaj, the bins coloured blue have $\DeltaE < 0$, while the bins coloured red have $\Delta E > 0$. The binning is used both at the reconstruction level $R$ and truth level $T$.}
\label{fig:unfolding:binning}
\end{figure}

Based on the selection criteria defined in Section~\ref{sec:event}, the \emph{optimised event rates} $N^{\text{opt}}(\theta_j, \Delta E)$ are extracted from the data in bins of the jet scattering angle, $\theta_j$, and the top--antitop energy difference, $\Delta E= E_{t} - E_{\bar{t}}$. Here $N^{\text{opt}}(\theta_j)$ is defined as
\begin{align*}
N^{\text{opt}}(\theta_j) & = N(\theta_j | y_{t\bar{t}j} > 0) + N(\pi - \theta_j | y_{t\bar{t}j} < 0)\,,
\end{align*}
in analogy with the optimised cross section $\sigma^{\text{opt}}(\theta_j)$ from Eq.~\eqref{eq:sigma-opt}. The selected data events are divided into three bins in $\theta_j$, $[0,\pi/4,3/5\pi,\pi]$, and four bins in $\Delta E$, $[-\infty,-100,0,100,\infty]\gev$. The choice of binning in $\theta_j$ optimizes the significance of the energy asymmetry in the central bin at particle level. The binning in $\Delta E$ was chosen to account for different migration behaviour for small and large energy differences, while at the same time maintaining a high event rate in each bin.
Each of the three $\thetaj$ bins is subdivided into four $\DeltaE$ bins for a total of 12 bins as illustrated in Figure~\ref{fig:unfolding:binning}. The chosen configuration shows the best response in linearity tests with different asymmetries in pseudo-data. To obtain $N^\text{opt}(\thetaj | \DeltaE > 0)$ and $N^\text{opt}(\thetaj | \DeltaE < 0)¸$ used in Eq.~\ref{eq:unfolding:AEN} the content of the corresponding pairs of bins are later summed.
 
The sample in the 12 bins measured at the detector level is corrected for detector effects (unfolded) to particle-level fiducial phase space defined in Section~\ref{sec:event}. The unfolding is performed with the Fully Bayesian Unfolding (FBU)~\cite{Choudalakis,TOPQ-2014-16} method, which applies Bayesian inference to obtain a full posterior probability density distribution of the unfolded asymmetry.
 
Given an observed spectrum $D\in \mathbb{N}^{N_\text{r}}$ with $N_\text{r}$ reconstructed bins, a `truth' spectrum $T\in \mathbb{R}^{N_\text{t}}$ with $N_\text{t}$ bins and a response matrix $\mathcal{M} \in \mathbb{R}^{N_\text{r}}\times\mathbb{R}^{N_\text{t}}$,  the posterior probability density of the true spectrum $T$ is calculated using Bayes' theorem:
\begin{equation}\nonumber
P(T|D,\mathcal{M})\propto \mathcal{L}(D|T,\mathcal{M})\pi(T),
\end{equation}
where $\mathcal{L}(D|T,\mathcal{M})$ is the likelihood function of $D$ for a given $T$ and $\mathcal{M}$, and $\pi(T)$ is the prior probability density for $T$. While the response matrix is estimated from the simulated sample
of $\ttbar$ events, a uniform prior probability density in all bins is
chosen as $\pi(T)$, such that equal probabilities are assigned to all $T$ spectra within a wide range between zero and twice the SM prediction.

The likelihood of observing data $D$ for various predictions of signal $R\in \mathbb{R}^{N_\text{r}}$ and background $B\in \mathbb{R}^{N_\text{r}}$ is given by the product of Poisson probabilities for each bin of the spectrum:
\begin{equation}\nonumber
\mathcal{L}(D|T,\mathcal{M},B)=\prod_{i=1}^{N_\text{r}}\frac{(r_i+b_i)^{d_i}}{d_i!}\text{e}^{-(r_i+b_i)},
\end{equation}
where $r_i$ and $b_i$ are the expected signal and background yields and $d_i$ is the observed data yield in bin $i$. The true distribution $T$ is related to the reconstructed signal distribution $R$ via the response matrix $\mathcal{M}$:
\begin{equation}
r_i(T,\mathcal{M})=\sum_{j=1}^{N_\text{t}}m_{ij}t_j, \qquad m_{ij}= \frac{\epsilon_{t_j} P(r_i|t_j)}{f_{\mathrm{acc},r_i}}
\label{eq:FBU_reco_def}
\end{equation}
with efficiency  $\epsilon_{t_j}$, migration $P(r_i|t_j)$ and acceptance $f_{\text{acc},r_i}$. The efficiency $\epsilon_{t_j}$  is the probability for an event produced in `truth' bin $t_j$ to be reconstructed in any bin, i.e.\ the combined detector acceptance, reconstruction efficiency and selection efficiency. The migration $P(r_i|t_j)$ is the probability for an event produced in the `truth' bin $j$ to be reconstructed in bin $i$, given that it is reconstructed in any bin. The acceptance $f_{\text{acc},r_i}$ is the fraction of events passing the detector-level selection in bin $i$ that also satisfy the fiducial phase-space selection at particle level.
 
\begin{figure}
\centering
\includegraphics[width=.5\textwidth]{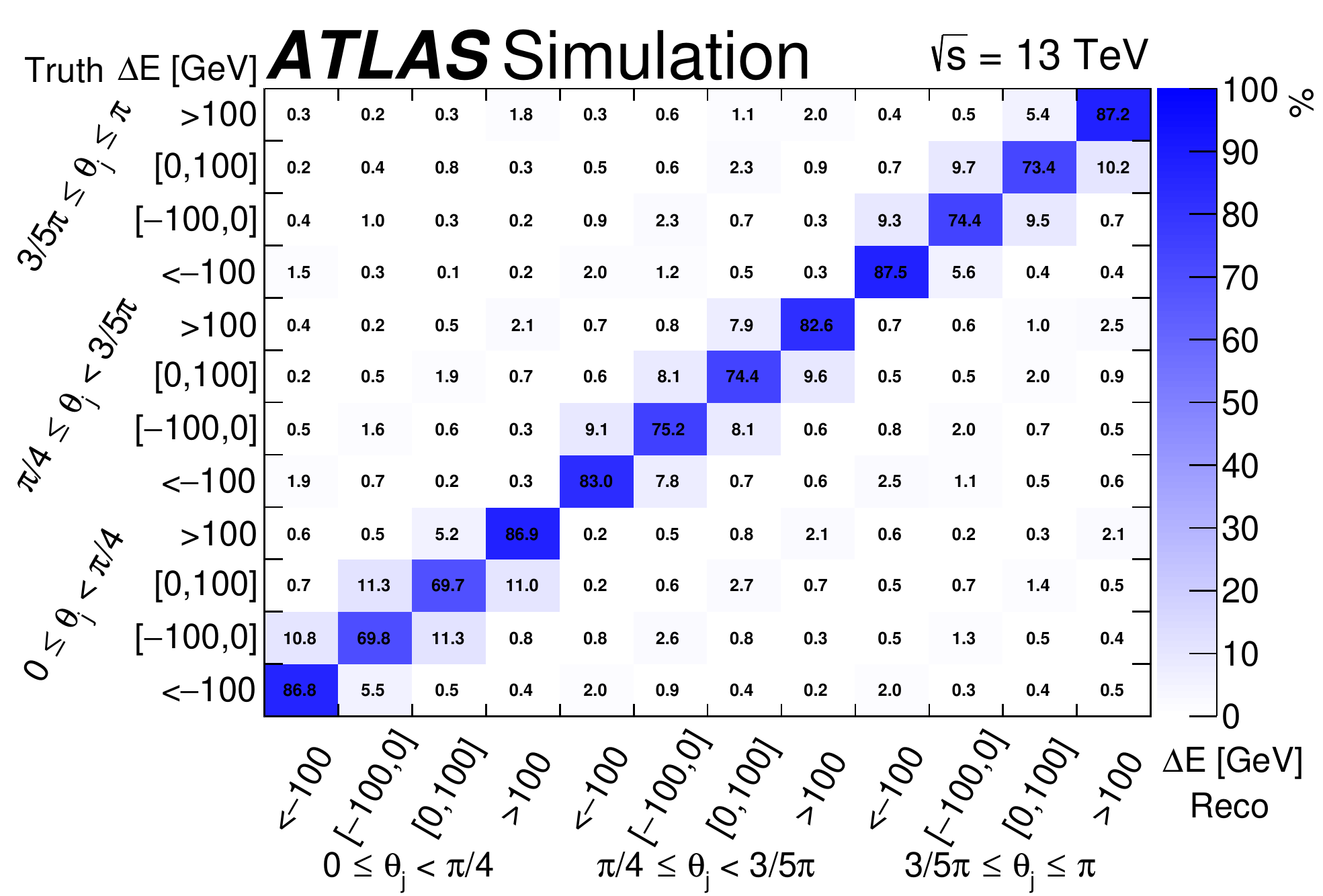}
\caption{Migration matrix, normalised to 100\% in each truth bin. This gives the probability of a reconstructed event created in truth bin $j$ to be reconstructed in bin $i$.}
\label{fig:unfolding:migmat}
\end{figure}

The migration matrix shown in Figure~\ref{fig:unfolding:migmat} is mostly diagonal, with the smallest diagonal entry being 69.7\%, indicating that the reconstruction is performing well. The average efficiencies and acceptances are $\epsilon = 34.4\%$ and $f_\text{acc} = 69.6\%$, respectively.
 
Systematic uncertainties are included in the FBU via marginalisation, i.e. by introducing $N_{\text{np}}$ nuisance parameters which are integrated out to obtain the marginal likelihood:
\begin{equation*}
\mathcal{L}(D|T)=\int \mathcal{L}(D|R(T;\pmb{\theta}_s),B(\pmb{\theta}_s,\pmb{\theta}_b))G(\pmb{\theta}_s)G(\pmb{\theta}_b)\text{d}\pmb{\theta}_s\text{d}\pmb{\theta}_b\text{,}
\end{equation*}
where $\pmb{\theta}_s$ are nuisance parameters for uncertainties in the modelling, object identification, reconstruction and calibration affecting both the signal prediction $R(T;\pmb{\theta}_s)$ and background prediction $B(\pmb{\theta}_s,\pmb{\theta}_b)$ and $\pmb{\theta}_b$ are the nuisance parameters for background normalisation uncertainties. The priors $G(\pmb{\theta_s})$ and $G(\pmb{\theta_b})$ are Gaussian and truncated Gaussian distributions, respectively, with $\mu=0$ and $\sigma=1$. The truncation is applied to avoid a non-physical negative number of events in the background estimate.
 
The posterior probability distribution $P(T|D)$ is determined by sampling the $(N_\text{t}+N_{\text{np}})$-dimensional parameter space, evaluating the likelihood $\mathcal{L}(D|T$,$\mathcal{M})\,\pi(T)$ for each point, and projecting the samples to the $T$-parameter space. The unfolded asymmetry $\ea(\thetaj)$ is computed from $P(T|D)$ as
\begin{equation}\nonumber
P(\ea(\theta_j)|D) = \int \delta(\ea(\theta_j)-\ea(\theta_j,T))P(T|D)\,\mathrm{d}T
\end{equation}
where $\ea(\theta_j,T)$ is defined as in Eq.~(\ref{eq:ea-opt:explicit}):
\begin{equation}
A_E(\thetaj,T) \equiv \frac{ N^\text{opt}(\thetaj | \DeltaE > 0) - N^\text{opt}(\thetaj | \DeltaE < 0)}{ N^\text{opt}(\thetaj | \DeltaE > 0) + N^\text{opt}(\thetaj | \DeltaE < 0) }
\label{eq:unfolding:AEN}
\end{equation}
where for $\Delta E > 0$ and $\Delta E < 0 $ the sum over the two corresponding bins is taken.
The estimate and variance of the energy asymmetry $\ea$ are given by the mean and variance of the posterior distribution. The marginalised posterior distributions of the nuisance parameters are given by projections on the corresponding one-dimensional nuisance parameter space.
The sampling is based on Markov Chain Monte Carlo methods ~\cite{Diaconis:2009} with the \textit{No-U-turn sampler} algorithm~\cite{Hoffman:2011} in four Markov chains each with 10\,000 sampling steps and 2500 tuning steps. The technical implementation uses the \textsc{PyFBU} code~\cite{PyFBU}, which uses the \textsc{PyMC3} package~\cite{PyMC3}.
 
The unfolding method was found to be robust against various reweightings of the input distributions, varying the energy asymmetry in a range between $-0.15$ and $0.15$, and with respect to the impact of SMEFT operators within current global limits. As expected, any difference between the unfolded and input asymmetries in these tests is covered by the uncertainties of the response matrix. The final result was also found to be stable with respect to the chosen flat prior. A Gaussian prior with the SM prediction as the mean and a 20\% uncertainty changes the central values by 2\%--6\% and results in approximately 4\% smaller uncertainties, supporting the argument that the flat prior is a conservative choice.

\section{Systematic uncertainties} 
\label{sec:syst}
Various experimental and theoretical systematic uncertainties affecting both the signal and background predictions are considered. A nuisance parameter with a Gaussian prior is assigned to each systematic uncertainty, unless otherwise specified, and their effect on the measurement is directly included in the unfolding procedure as described in Section~\ref{sec:unfolding}. The smoothing and symmetrisation procedures are described in Section~\ref{sec:systematics:smoothing_symmetrisation}.
 
\subsection{Experimental uncertainties}
The uncertainty in the combined 2015--2018 integrated luminosity is 1.7\% \cite{ATLAS-CONF-2019-021}, obtained using the LUCID-2 detector \cite{LUCID2} for the primary luminosity measurements. This uncertainty affects only the backgrounds estimated from simulation, since the luminosity uncertainty in the signal cancels out when measuring the energy asymmetry.
A reweighting procedure based on the mean number of interactions is used to correct  the pile-up profile in MC simulation to match the data.
A systematic uncertainty is assigned to the weights.
Lepton ($e,\mu$) identification, reconstruction, isolation and trigger performance~\cite{TRIG-2018-05,TRIG-2018-01} can differ slightly between data and simulation such that scale factors are applied to correct for these differences. The scale factors were obtained by a tag-and-probe method in $Z$ boson and $J/\psi$ decays~\cite{PERF-2017-01,MUON-2018-03}. Systematic uncertainties are assigned to each of these scale factors.  The lepton momentum scale and resolution are corrected for the differences between data and simulation, and the systematic uncertainties for these corrections are taken into account.
 
For jets, the jet energy scale and resolution uncertainties are applied to both the \smallR and \largeR jets, taking correlations into account~\cite{PERF-2016-04,PERF-2014-02,JETM-2018-02}. Further uncertainties are considered for the JVT~\cite{PERF-2014-03}, the jet mass calibration~\cite{JETM-2018-02}, and the top-tagging scale factors~\cite{JETM-2018-03,ATL-PHYS-PUB-2020-017} for tagged and mistagged jets.
 
Eigenvector variations for $b$-, $c$- and light-jet $b$-tagging efficiencies~\cite{PERF-2016-05} are applied to the flavour-tagging scale factors, together with the uncertainty on the $\pT$ extrapolation to the regions in which they could not be determined directly.
The uncertainties of all the $\met$ components associated with reconstructed objects are propagated in a fully correlated way. Additional uncertainties in the scale and resolution of the soft-track components are considered \cite{PERF-2016-07}.
 
Three kinds of uncertainties are considered for the non-prompt- and fake-lepton backgrounds: the uncertainty in the prediction, due to the limited number of data events and large applied weights, which is uncorrelated for each unfolded bin; the statistical uncertainties of the real- and fake-lepton efficiencies for electrons and muons (four parameters); and a 50\% normalisation uncertainty for this background~\cite{TOPQ-2018-15}.
 
\subsection{Theoretical uncertainties}
The relative uncertainties in the inclusive cross sections of backgrounds estimated from simulation are 5.3\% for single-top~\cite{Aliev:2010zk,Kant:2014oha} and 5\% for $W$+jets~\cite{STDM-2015-03}. A conservative 50\% normalisation uncertainty is applied for the remaining $Z$+jets, diboson, $t\bar{t}V$ and $t\bar{t}H$ background samples \cite{TOPQ-2018-15}, which together represent only 3\% of the total expected event yield.
In data/MC comparison plots, a relative uncertainty of 6.1\% is applied to the $t\bar{t}$ cross section~\cite{Czakon:2011xx}. In the unfolding the amount of signal is unconstrained and not affected by this systematic uncertainty.
 
The effect of signal modelling uncertainties on the signal distribution $R$ is estimated by varying the response matrix in Eq.~(\ref{eq:FBU_reco_def}) for the fixed nominal `truth' distribution, since the unfolding procedure depends only on the response matrix and is performed without any assumption about the underlying `truth' distribution.
 
Signal modelling uncertainties account for the following possible sources of systematic uncertainty:
\begin{itemize}
\item parton shower and hadronisation model, estimated as the symmetrised difference between \pythia and \herwigseven, both interfaced to \powhegbox;
\item matching between matrix element and parton shower, evaluated from the difference between the \powhegbox and the \mgamcshort predictions, both interfaced to \pythia, and from variations of the \hdamp parameter;
\item initial-state radiation, derived from varied weights in the matrix element and in the parton shower obtained in simulations with $\muR$ and $\muF$ values different from the nominal one;
\item final-state radiation, based on the variation of parton shower weights corresponding to different renormalisation scales;
\item top-quark mass used in the simulation.
\end{itemize}
The samples used to estimate these uncertainties are described in Section~\ref{sec:samples:ttsig}.
 
The uncertainties in parton distribution functions are evaluated using a set of 30 nuisance parameters described in the PDF4LHC15 prescription~\cite{Rojo:2016ymp}. The PDF variations are obtained using alternative MC generator weights stored within the nominal \textsc{Powheg\,+\,Pythia\,8} \ttbar sample.
The uncertainty from the limited size of the MC signal sample is estimated using a bootstrapping method outlined in Ref.~\cite{Bohm:169869} from the covariance matrix of the ensemble of repeated unfolding results with varied response matrices.
 
The single-top systematic uncertainties are decorrelated between the s-, t- and $Wt$-channel, except for the PDF uncertainties, which are fully correlated across the different channels and with \ttbar. Both the differences in shape and normalisation are taken into account, except for the PDF uncertainties, since their effect on the normalisation is already included in the cross-section uncertainty.
The modelling uncertainties are the same ones as considered for \ttbar and are estimated with the same methods. In addition, the difference between the predictions for the $Wt$-channel obtained with diagram subtraction or removal is considered as a systematic uncertainty~\cite{Frixione:2008yi}.
 
For the \Wjets background, modelling uncertainties are estimated by varying the renormalisation and factorisation scales up and down by a factor of 2, varying the matrix element to parton shower matching scale up and down by factors of 1.5  and 1.33, respectively, and varying the parton shower resummation scale up and down by a factor of 4, in each case relative to scale(s) for the nominal sample described in Section~\ref{sec:background:prompt} (see \Bibref{ATL-PHYS-PUB-2017-007} for further details).
The uncertainties in the parton distribution functions are evaluated using a set of 100 PDF variations included in the \textsc{NNPDF3.0} description \cite{Ball:2014uwa}, obtained using alternative MC generator weights stored within the nominal \textsc{Sherpa\,2.2.1} \Wjets samples. To account for the systematic uncertainty in the fraction of \Wjets events containing jets originating from b- or c-quarks, these events are filtered into separate samples and their individual normalisation is varied up and down by 30\%~\cite{ATL-PHYS-PUB-2017-006}.
 
\subsection{Smoothing and symmetrisation}
\label{sec:systematics:smoothing_symmetrisation}
 
In order to mitigate effects due to statistical fluctuations in the MC samples used to estimate systematic uncertainties, a bootstrapping method~\cite{Bohm:169869} is used to remove statistically insignificant effects in the \ttbar, single-top and \Wjets modelling uncertainties.
A variation in a detector-level bin corresponding to one standard deviation of the systematic uncertainty is considered to be statistically significant if it is more than twice as large as the statistical uncertainty of the expected yield in the bin. If at least one of the four $\Delta E$ bins in a $\theta_j$ interval has a significant variation, all four are left unchanged and no smoothing is applied. Same-sign $\Delta E$ bins are merged and kept if the variation is statistically significant after adding them up, but no variation in the single bins is itself statistically significant. Otherwise, all four bins are combined and the same significance criterion is applied to the combination. If it is significant, all four bins are set to this combined value, effectively dropping the shape of the systematic uncertainty, but keeping its normalisation. If it is still not significant, the systematic uncertainty is neglected in the unfolding.
 
Since a nuisance parameter with a Gaussian prior is assigned to each of them, the uncertainties are symmetrised in each bin of the reconstruction-level distribution. For one-sided systematic uncertainties, the difference between the nominal yield and the varied yield is used as the standard deviation of the Gaussian prior. For two-sided systematic uncertainties, the average absolute difference between the nominal yield and the upper- and lower-side varied yields is used as the standard deviation.

\section{Measurement results} 
\label{sec:SMresult}
\begin{figure}
\centering
\includegraphics[width=.5\textwidth]{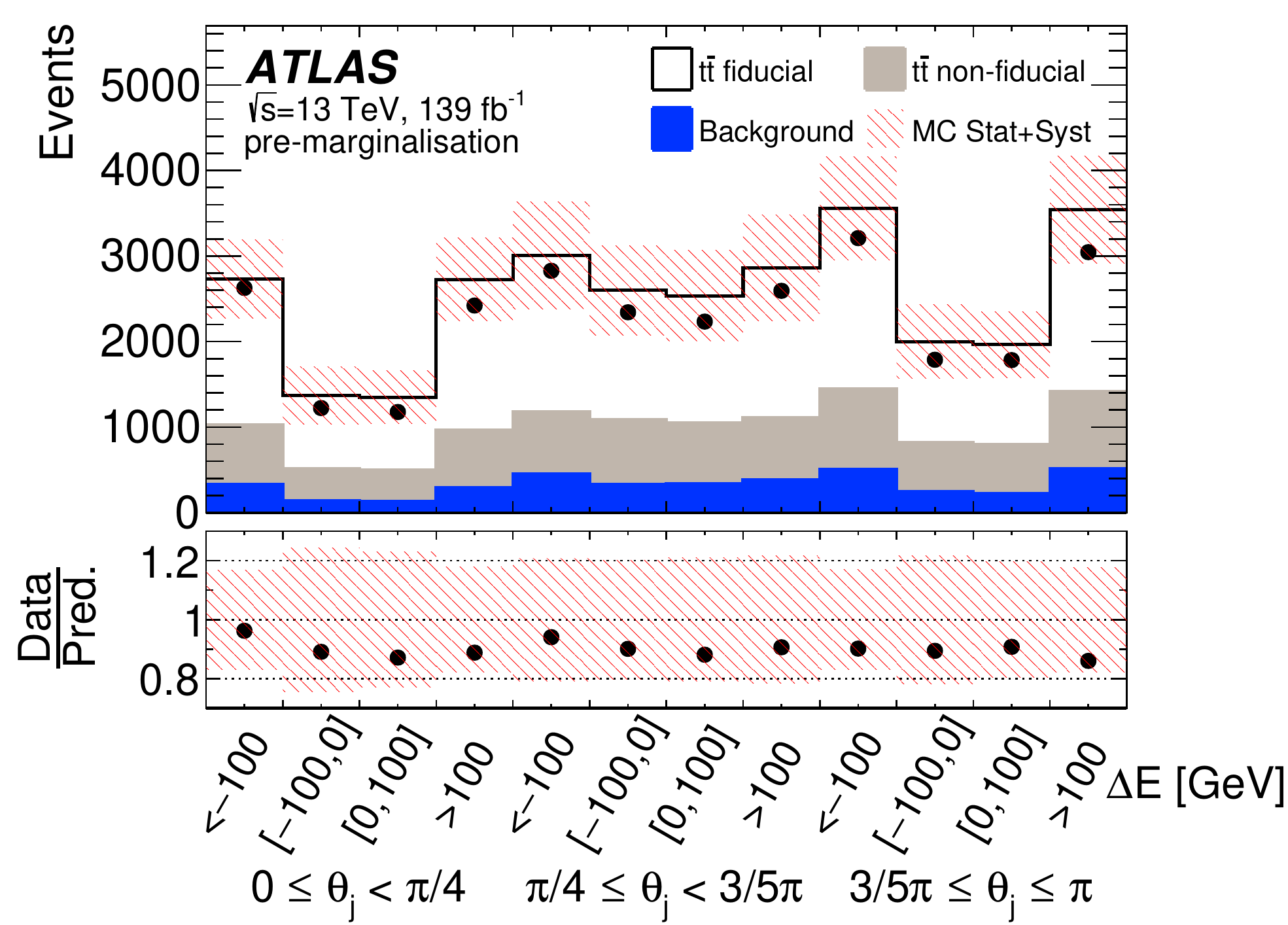}
\caption{Comparison of data to the prior (pre-marginalisation) prediction for the $\DeltaE$ vs. $\thetaj$ distribution used in the unfolding. The 3 differential bins in $\thetaj$ and 4 $\DeltaE$ bins are concatenated to give 12 bins as shown in Figure~\ref{fig:unfolding:binning}. The shaded red bands correspond to the total uncertainty of the prediction. }
\label{fig:thetajopt_vs_deltaE}
\end{figure}

The distribution of the data events at detector level binned in \DeltaE and \thetaj is compared with the SM expectation in Figure~\ref{fig:thetajopt_vs_deltaE}.
The data are consistent with the predictions within uncertainties. This distribution is the input to the unfolding procedure. With the same number of detector-level bins as `truth' level bins, the posterior prediction agrees with the data distribution and is thus not shown.
 
\begin{figure}
\centering
\includegraphics[width=.32\textwidth]{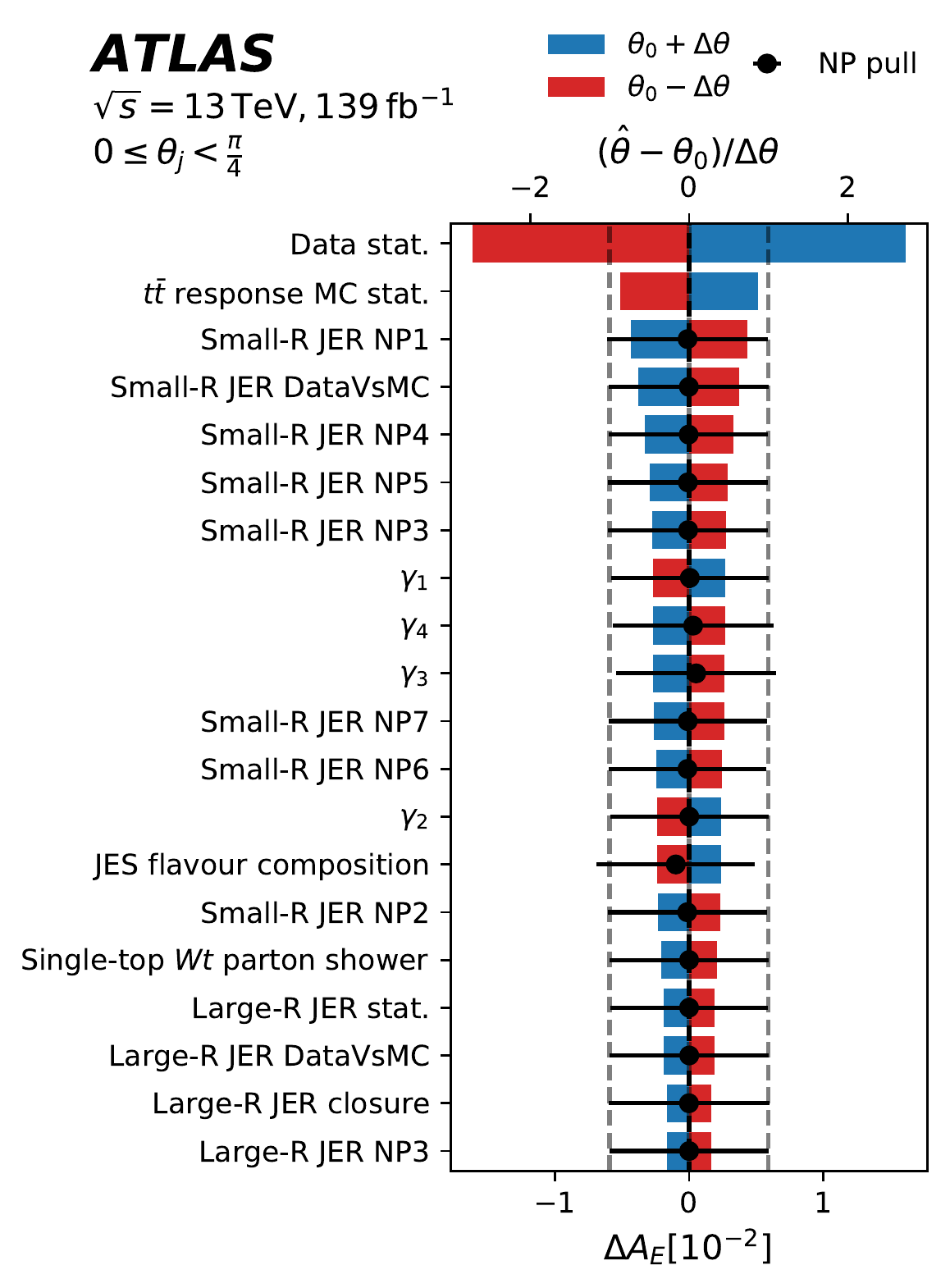}
\includegraphics[width=.32\textwidth]{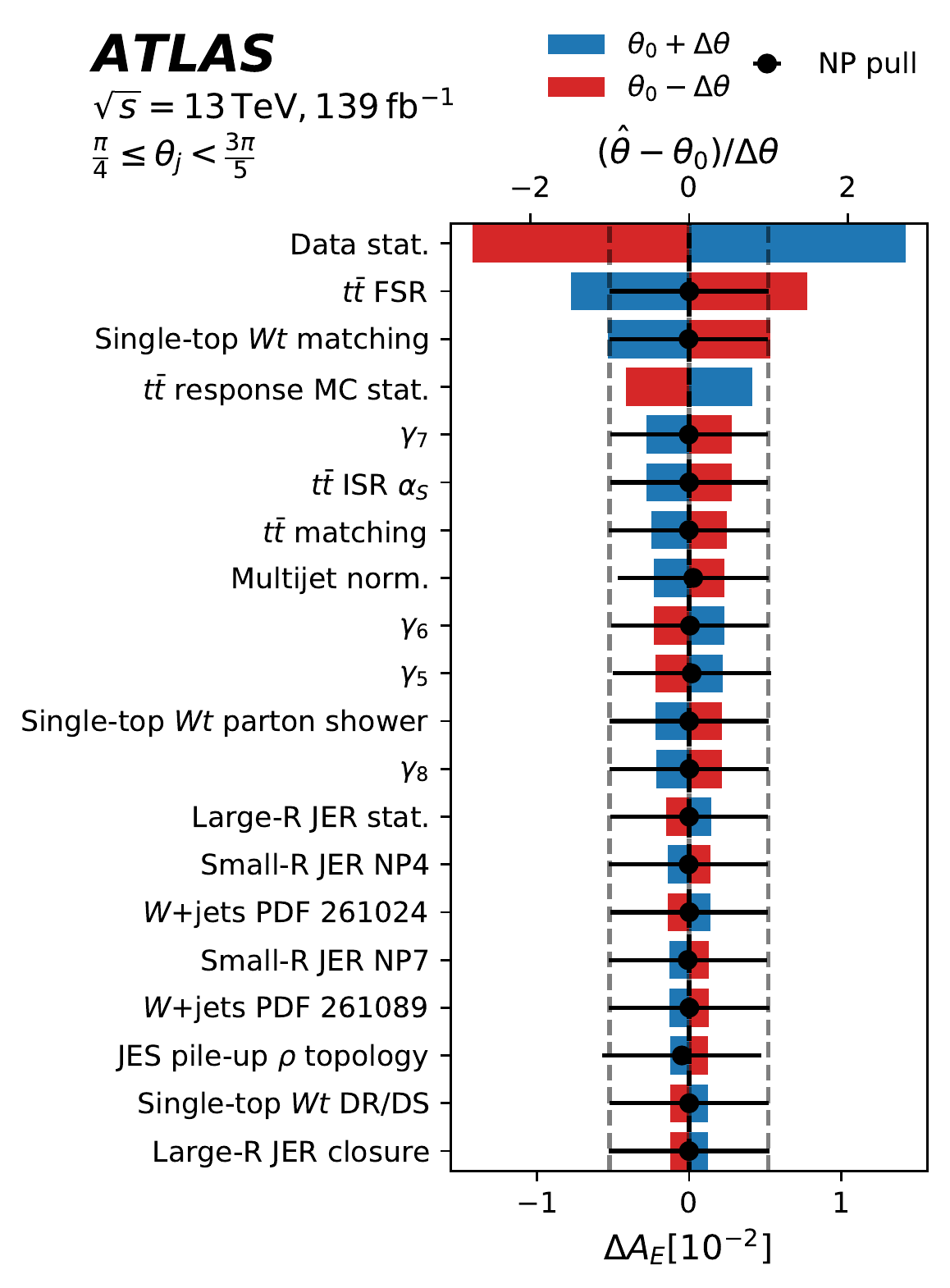}
\includegraphics[width=.34\textwidth]{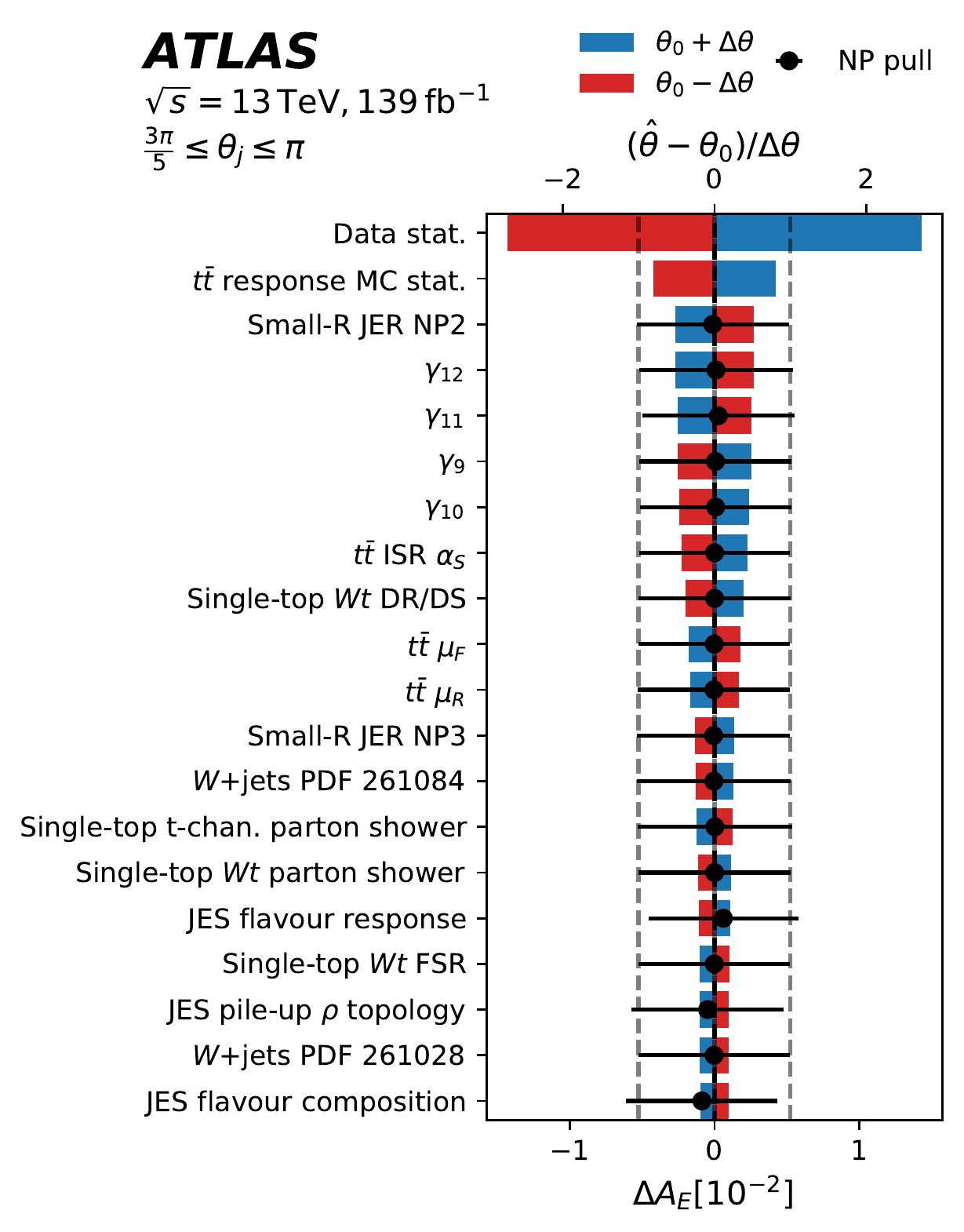}\\
\caption{Ranking of systematic uncertainties in the three \thetaj bins.
The black dots and error bars (upper scale) denote the mean and standard deviation of the posterior distribution of the nuisance parameter, normalised to its pre-marginalisation standard deviation. The blue and red boxes (lower scale) show the impact on the energy asymmetry from a $1\sigma$ variation in its influencing nuisance parameter.
The $\gamma$ variations denote the statistical uncertainties of the background predictions in the corresponding bin of the $\DeltaE$ vs $\thetaj$ distribution. The numbers appended to the \Wjets PDF variations denote the corresponding \textsc{NNPDF3.0}~\cite{Ball:2014uwa} PDF sets.}
\label{fig:impact}
\end{figure}

To determine the impact of each systematic uncertainty, a method of ranking is employed in Figure~\ref{fig:impact}. For each bin, the effects of $1 \sigma$ variations in its influencing nuisance parameters are calculated. These effects are then used to rank each systematic uncertainty for the three \thetaj bins individually. The effects are extracted from the samples of the posterior distribution with $\sigma_i^{(j)} = c_{ij}/\sqrt{c_{jj}}$ being the estimated shift of bin $i$ due to a shift of nuisance parameter $j$, where $c_{ij}$ denotes the covariance of the posterior distributions.
The impact of systematic uncertainties is largest in the central bin due to the $t\bar{t}$ FSR variation and the choice of matching scheme in the single-top $Wt$ channel being significant compared to the statistical precision of the MC samples, whereas the impact is negligible in the outer bins and therefore ignored in accord with the smoothing procedure described in Section~\ref{sec:systematics:smoothing_symmetrisation}. The different size of the systematic effects is explained by the difference in the relative populations of the $\Delta E$ bins in each angular interval, causing bin-to-bin migrations produced by the systematic variations to affect the asymmetry most strongly in the central $\theta_j$ bin.
The statistical uncertainties displayed in Figure~\ref{fig:impact} are given by the uncertainty from unfolding with all nuisance parameters fixed to their post-marginalisation values. The MC statistical uncertainty of the response matrix is derived with the bootstrapping method described in Section~\ref{sec:syst}. The impact of systematic uncertainties by category is calculated with Gaussian error propagation and shown in Table~\ref{tab:results:impact}. The measurement is dominated by the data statistical uncertainty; the largest systematic uncertainties are from the jet energy resolution, the $\ttbar$ modelling and the statistical uncertainty of the multijet estimate in the first, second and third bins, in that order. The first $\thetaj$ bin preferentially selects jets closer to the beamline as compared to the other bins; the jet energy resolution uncertainty is therefore larger, as this uncertainty increases with the absolute value of $\eta$ within the analysis fiducial volume~\cite{JETM-2018-05}.
 
\begin{figure}
\centering
\includegraphics[width=.5\textwidth]{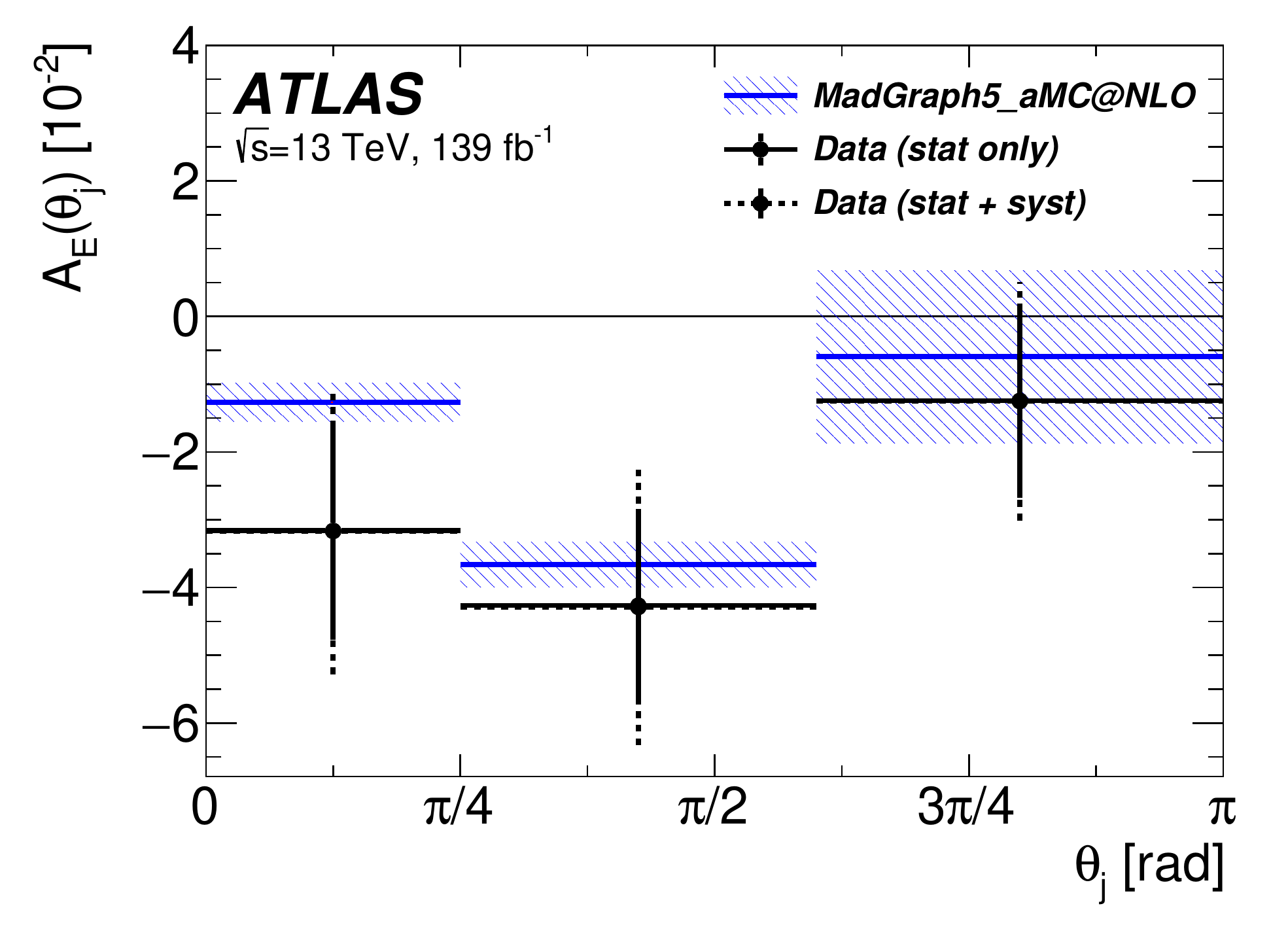}
\caption{Data measurements (black points with errors) and predictions (blue lines) of the energy asymmetry in three bins of the jet angle $\theta_j$. The SM prediction was obtained from simulations of $\ttj$ events with \mgamc+~\pythia at NLO in QCD for $\ttj$ + PS, including MC statistical and scale uncertainties (blue shaded bands) as described in Section~\ref{sec:samples:ttsig}.}
\label{fig:results:asymmetries}
\end{figure}

The measured \ea is obtained from the unfolded \DeltaE distributions in the three \thetaj bins and is shown in Table~\ref{tab:results} and in Figure~\ref{fig:results:asymmetries} together with the SM prediction including MC statistical and scale uncertainties obtained as described in Section~\ref{sec:samples:ttsig}. In the highest $\theta_j$ bin the uncertainty in the SM prediction is similar to the expected sensitivity of the measurement and dominated by scale uncertainties. The measured energy asymmetry is in good agreement with the SM prediction with a combined $p$-value of $0.80$. In the central bin, the measured (expected) energy asymmetry differs from zero by 2.1 (1.9) standard deviations.
While the systematic uncertainties of the total cross section dominate for the \DeltaE and \thetaj distribution in Figure~\ref{fig:thetajopt_vs_deltaE}, these uncertainties cancel out in the \ea measurement. The correlations between \ea values measured in pairs \thetaj bins are always smaller than 5\% in magnitude.
 
\begin{table}
\caption{Impact of systematic uncertainties by category.
The impact of the individual uncertainties is estimated from the posterior distribution as the effect on a bin $i$ of a $1 \sigma$ variation in its influencing nuisance parameter $j$, $\sigma_i^{(j)} = c_{ij}/\sqrt{c_{jj}}$. The impact of a group of uncertainties is calculated with Gaussian error propagation, taking into account correlations between uncertainties. The uncertainties in the luminosity and the cross sections of \Zjets, \VV, \ttV and \ttH backgrounds are found to be negligible and are not shown in the table.
}
\vspace{10pt}
\label{tab:results:impact}
\centering
\begin{tabular}{l|ccc}
\multirow{2}{*}{Scenario} & \multicolumn{3}{c}{$\Delta A_E~[10^{-2}]$} \\
& $\thetajfirst$ & $\thetajcenter$ &  $\thetajlast$ \\
\toprule
Data statistical uncertainty &  $1.60$ & $1.40$ & $1.40$\\
\midrule
\ttbar modelling &  $0.08$ & $0.87$ & $0.34$\\
\ttbar response MC statistics & $0.51$ & $0.42$ & $0.42$\\
\Wjets modelling and PDF& $0.29$ & $0.49$ & $0.42$\\
\singletop modelling & $0.28$ & $0.60$ & $0.29$\\
\ttbar and single-top PDF & $0.08$ & $0.10$ & $0.07$\\
Multijet & $0.53$ & $0.54$ & $0.51$\\
Jet energy resolution & $0.98$ & $0.40$ & $0.36$\\
Other detector uncertainties & $0.42$ & $0.43$ & $0.30$\\
\midrule
Total & $2.10$ & $2.00$ & $1.80$\\
\bottomrule
\end{tabular}
\end{table}
 
\begin{table}
\caption{The measurement (`Data') of the energy asymmetry in three bins of the jet angle $\theta_j$ is compared with the `SM prediction' obtained from simulations of $\ttj$ events with \mgamc+~\pythia at NLO in QCD for $\ttj$ + PS, including MC statistical and scale uncertainties as described in Section~\ref{sec:samples:ttsig}. The table also shows the expected measurement values and uncertainties in the case of observing exactly the `SM prediction' (labelled as ’SM expectation’).}
\vspace{10pt}
\label{tab:results}
\centering
\begin{tabular}{l|ccc}
\multirow{2}{*}{Scenario} & \multicolumn{3}{c}{$A_E \pm \Delta A_E \, [10^{-2}]$} \\
& $0 \le \theta_j \le \frac{\pi}{4}$ & $\frac{\pi}{4} \le \theta_j \le \frac{3\pi}{5}$ &  $\frac{3\pi}{5} \le \theta_j \le \pi$ \\
\toprule
Data & $ -3.2\pm 2.1$ & $-4.3\pm 2.0$ & $-1.3\pm 1.8$\\
SM prediction (\mgamc) & $-1.3\pm0.3$ & $-3.7\pm0.3$ & $-0.6\pm1.3$\\
SM expectation & $-1.3\pm2.1$ & $-3.7\pm2.0$ & $-0.6\pm1.6$\\
\bottomrule
\end{tabular}
\end{table}

\section{SMEFT interpretation} 
\label{sec:smeft}
The sensitivity of the energy asymmetry to new physics above a scale $\Lambda$ is investigated in the framework of the Standard Model effective field theory (SMEFT). To this end, the SM Lagrangian $\mathcal{L}_\text{SM}$ is extended by a series of local operators $O_k$,
\begin{equation}\label{eq:leff}
\mathcal{L}_\text{SMEFT} = \mathcal{L}_\text{SM} + \sum_k \frac{C_k}{\Lambda^2}\,O_k + \dots\,.
\end{equation}
The sum runs over all SM gauge-invariant operators of mass dimension six. The ellipsis indicates higher-dimensional operators in the SMEFT expansion, which are neglected in this analysis. The analysis also assumes $CP$ invariance, which implies that all Wilson coefficients $C_k$ are real. Setting all the Wilson coefficients to zero corresponds to the SM. The cut-off scale of this effective theory is set to $\Lambda = 1\tev$ throughout the analysis.
 
In the Warsaw basis \cite{Grzadkowski:2010es}, 15 dimension-six operators affect $t\bar{t}j$ production at tree level \cite{Basan:2020btr}. The energy asymmetry is particularly sensitive to the chirality and the colour charges of the involved quark fields. To demonstrate this sensitivity, the focus of this numerical analysis is on six selected four-quark operators \cite{AguilarSaavedra:2018nen}
\begin{align}\label{eq:operators}
O_{Qq}^{1,8} & = (\bar{Q}_\text{L}\gamma_\mu T^A Q_\text{L})(\bar{q}_\text{L}\gamma^\mu T^A q_\text{L}) & O_{Qq}^{1,1} & = (\bar{Q}_\text{L}\gamma_\mu Q_\text{L})(\bar{q}_\text{L}\gamma^\mu q_\text{L}) \nonumber \\
O_{tu}^8 & = (\bar{t}_\text{R}\gamma_\mu T^A t_\text{R})(\bar{u}_\text{R}\gamma^\mu T^A u_\text{R}) & O_{tu}^1 & = (\bar{t}_\text{R}\gamma_\mu t_\text{R})(\bar{u}_\text{R}\gamma^\mu u_\text{R}) \nonumber \\
O_{tq}^8 & = (\bar{q}_\text{L}\gamma_\mu T^A q_\text{L})(\bar{t}_\text{R}\gamma^\mu T^A t_\text{R}) & O_{tq}^1 & = (\bar{q}_\text{L}\gamma_\mu q_\text{L})(\bar{t}_\text{R}\gamma^\mu t_\text{R})\,.
\end{align}
Left-handed quark doublets of the first two generations and the third generation are denoted by
\begin{align}\nonumber
q_\text{L} & = (u_\text{L},d_\text{L})^\top\,,\ (c_\text{L},s_\text{L})^\top\,,\qquad Q_\text{L} = (t_\text{L},b_\text{L})^\top\,,
\end{align}
reflecting an assumed flavour symmetry among couplings of the first and second quark generations. These six operators feature different chiral structures (LL, RR, LR) and colour structures (singlet (1) and octet (8)), thus covering the relevant properties that can be tested with the energy asymmetry. Operators with right-handed down-type quarks $d_\text{R}$ would affect the energy asymmetry very similarly to those with up-type quarks and are therefore not included in this analysis. Up- and down-quark operators can be resolved for instance with $t\bar{t}Z$ or $t\bar t W$ observables, which are sensitive to the weak isospin of the quarks~\cite{Brivio:2019ius}.
 
The Wilson coefficients in the effective Lagrangian in Eq.~\eqref{eq:leff} parameterize possible effects of virtual new heavy particles in the cross section $\sigma$ and in asymmetry observables $A$ according to
\begin{align}\label{eq:obs-smeft}
\sigma & = \sigma_\text{S}^{\text{SM}} + \sum_k\,C_k\sigma_\text{S}^k + \sum_{k\le l} C_k C_l\,\sigma_\text{S}^{kl}\,,\qquad A = \frac{\sigma_\text{A}}{\sigma_\text{S}} =
\frac{ \sigma_\text{A}^{\text{SM}} + \sum_k C_k  \sigma_\text{A}^{k}
+ \sum_{k\le l} C_k C_l \, \sigma_\text{A}^{kl}}{ \sigma_\text{S}^{\text{SM}} + \sum_k C_k \sigma_\text{S}^{k} +  \sum_{k\le l} C_k C_l\, \sigma_\text{S}^{kl}}\,.
\end{align}
The cross sections $\sigma_\text{S,A}^{\text{SM}}$ denote the SM contributions, while $\sigma_\text{S,A}^k$ and $\sigma_\text{S,A}^{kl}$ correspond to the interference of a SMEFT amplitude with the SM amplitude at $\mathcal{O}(\Lambda^{-2})$ and the pure SMEFT amplitudes at $\mathcal{O}(\Lambda^{-4})$, respectively. For the energy asymmetry the charge-symmetric and charge-asymmetric cross sections $\sigma_\text{S}$ and $\sigma_\text{A}$ are defined as
\begin{align}
\sigma_\text{S} & = \sigma^\text{opt}(\thetaj | \DeltaE > 0) + \sigma^\text{opt}(\thetaj | \DeltaE < 0) \nonumber \\
\sigma_\text{A} & = \sigma^\text{opt}(\thetaj | \DeltaE > 0) - \sigma^\text{opt}(\thetaj | \DeltaE < 0)\,, \nonumber
\end{align}
with $\sigma^\text{opt}$ defined as in Eq.~(\ref{eq:sigma-opt}).
The SM prediction and SMEFT contributions are computed in $\ttj$ production at NLO and LO in QCD, respectively, as described in Section~\ref{sec:samples:ttsig}. Relative to the SM predictions $\sigma_\text{S,A}^{\text{SM}}$, the four-quark operator contributions $\sigma_\text{S,A}^k$ and $\sigma_\text{S,A}^{kl}$ grow at high energies as $(m_{t\bar t}/2m_t)^2$ and $(m_{t\bar t}/2m_t)^4$, respectively~\cite{Brivio:2019ius}. In this analysis, the energy asymmetry is defined in the boosted regime, which probes invariant masses in the range $600 \lesssim m_{t\bar t} \lesssim 1200~\GeV$, well above the $t\bar t$ threshold. The energy enhancement partially cancels out in the normalised asymmetry, but is still present due to the growth of the charge-asymmetric cross section $\sigma_\text{A}$ in $\Delta E$ at high energies.
 
\begin{figure}[t!]
\centering
\subfloat[]{\includegraphics[width=.48\textwidth]{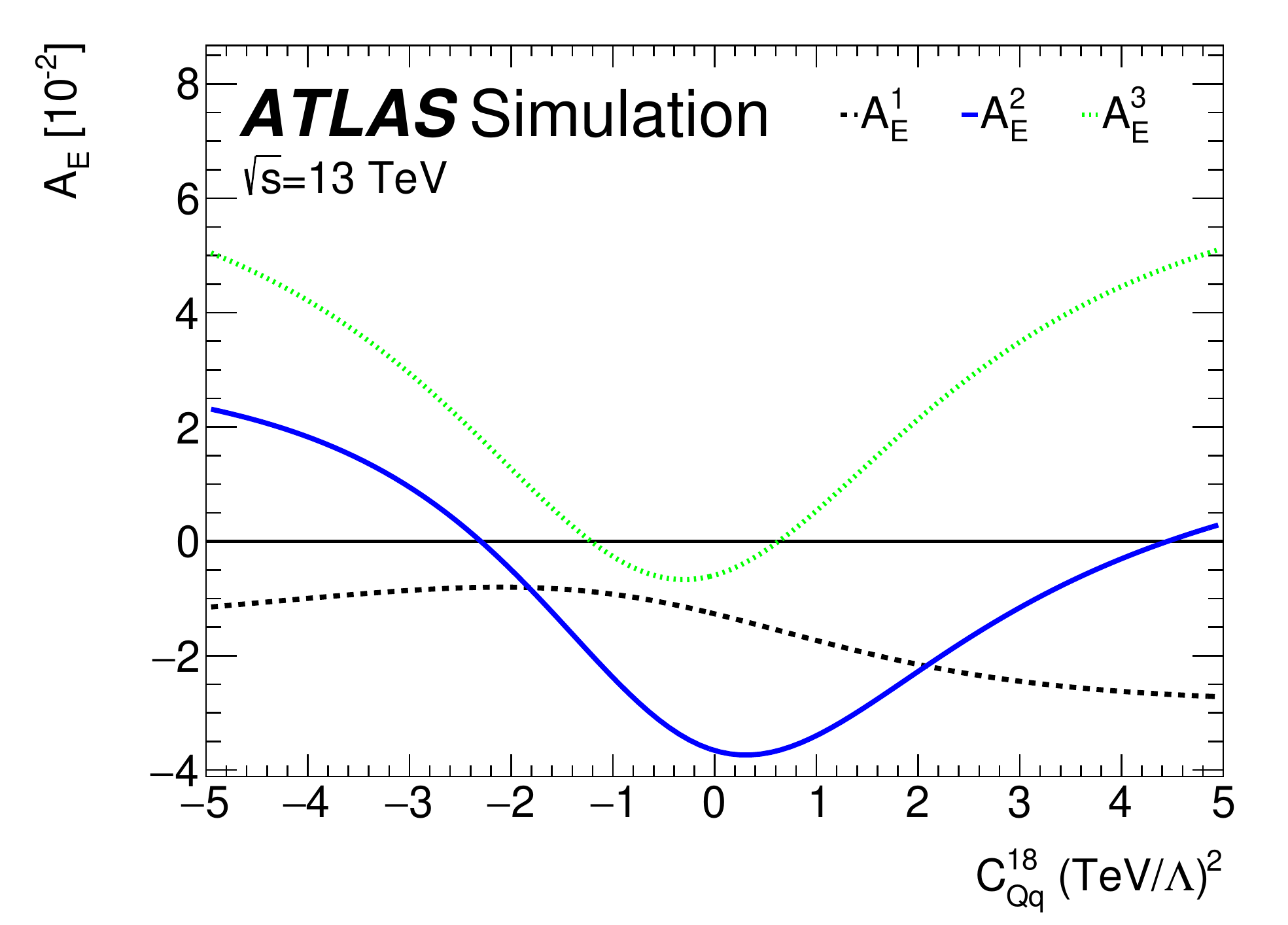}}\hspace*{0.5cm}
\subfloat[]{  \includegraphics[width=.48\textwidth]{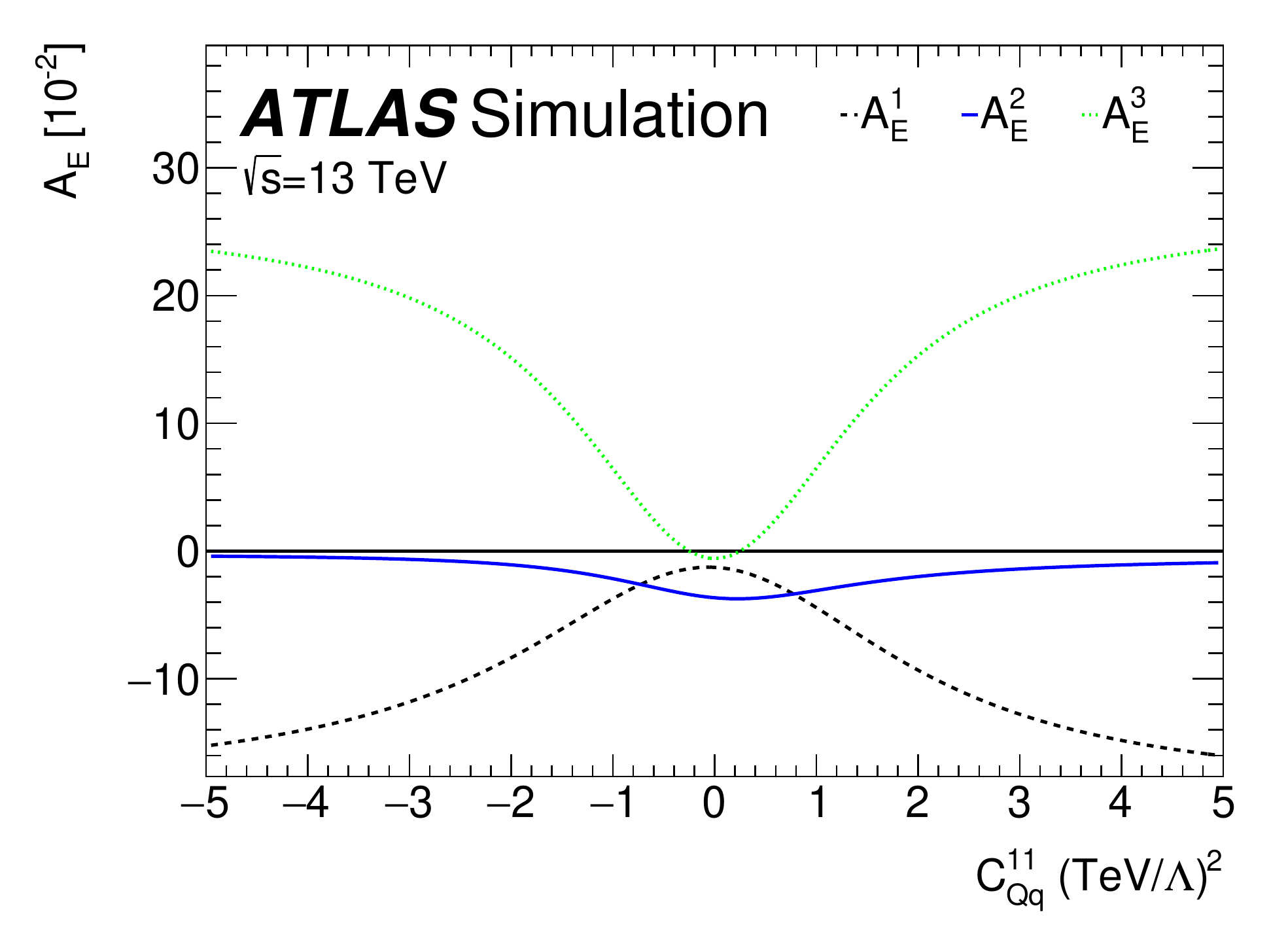}}\\
\subfloat[]{ \includegraphics[width=.48\textwidth]{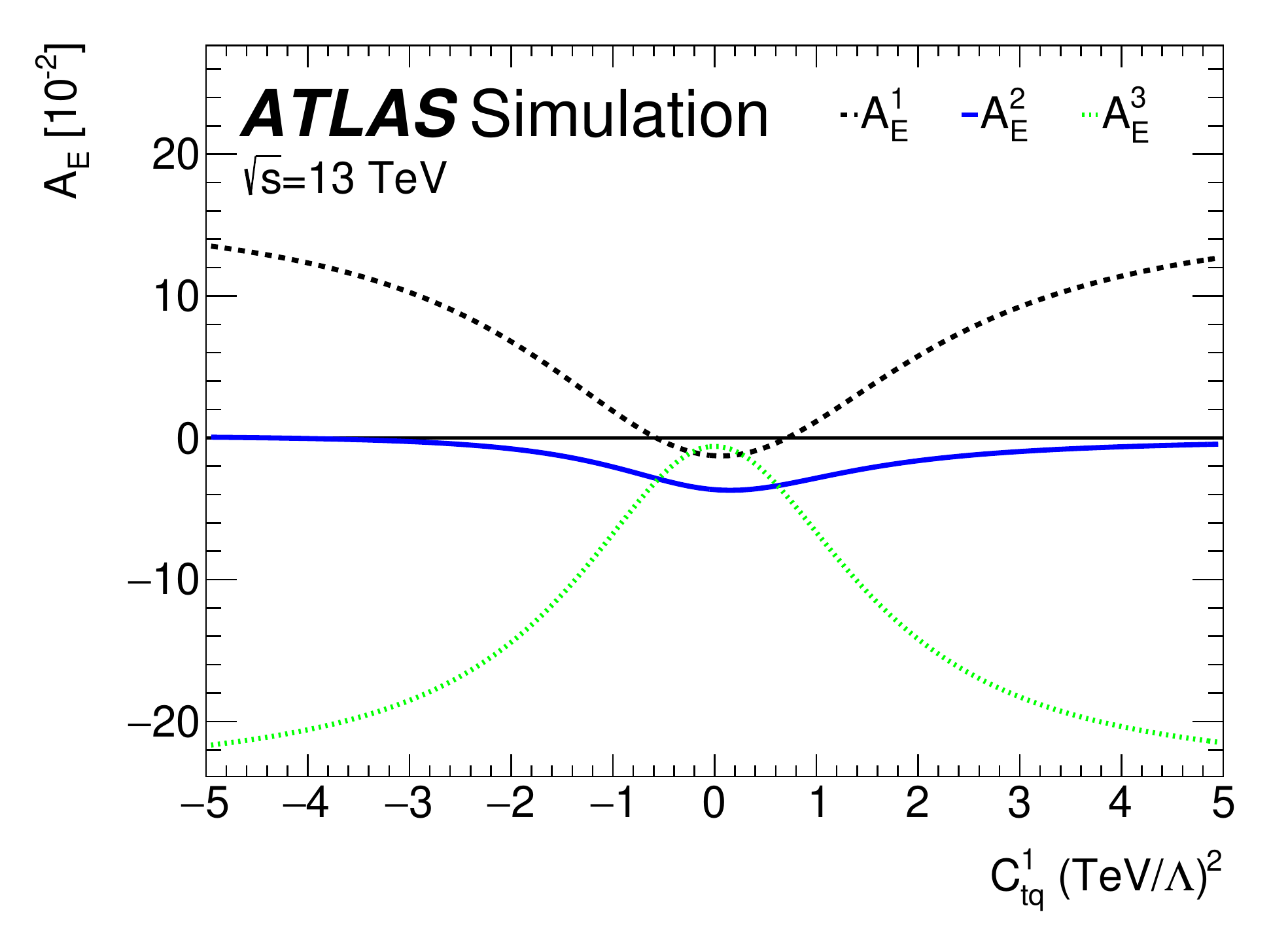}}\hspace*{0.5cm}
\subfloat[]{  \includegraphics[width=.48\textwidth]{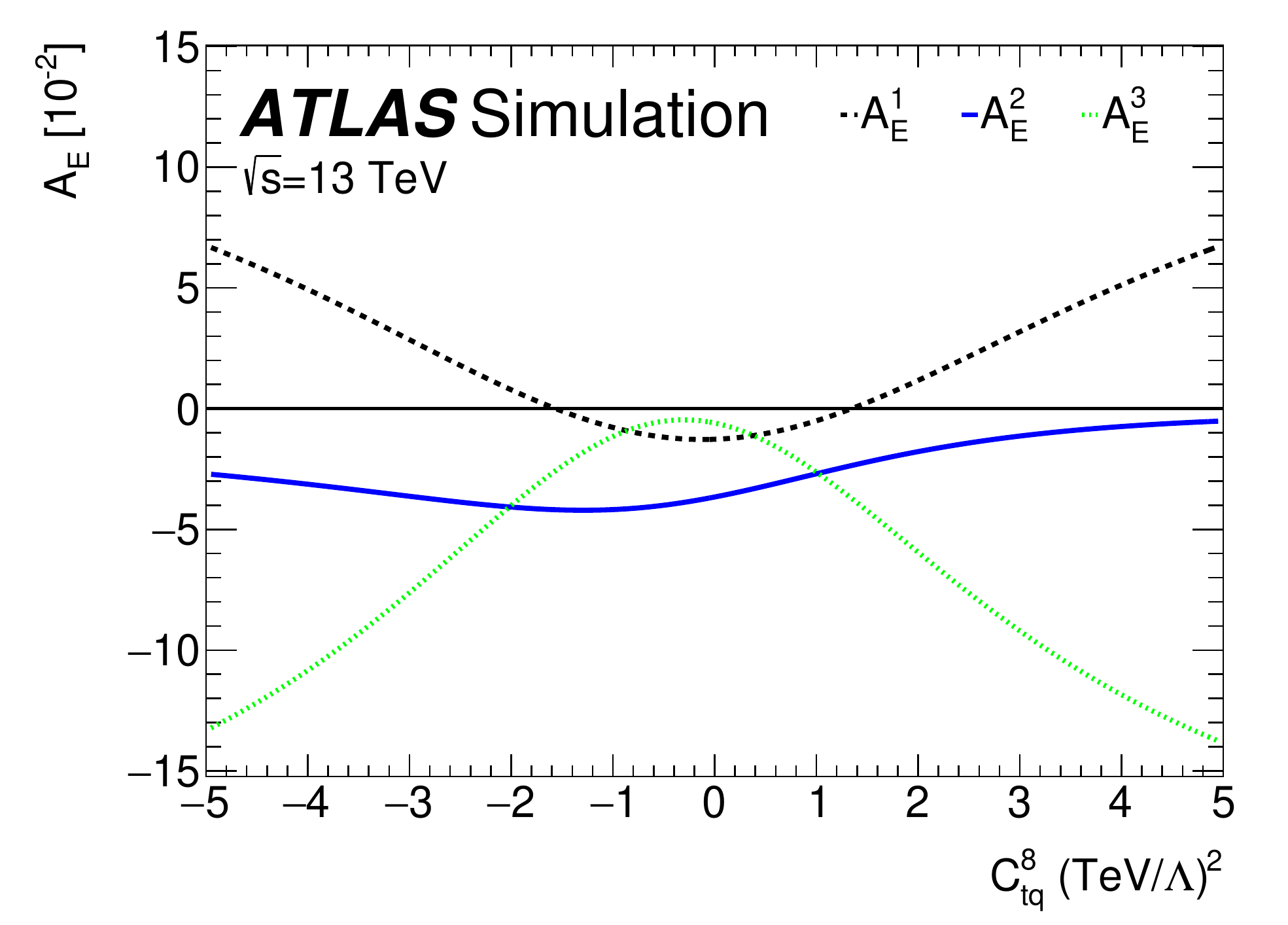}}\\
\subfloat[]{ \includegraphics[width=.48\textwidth]{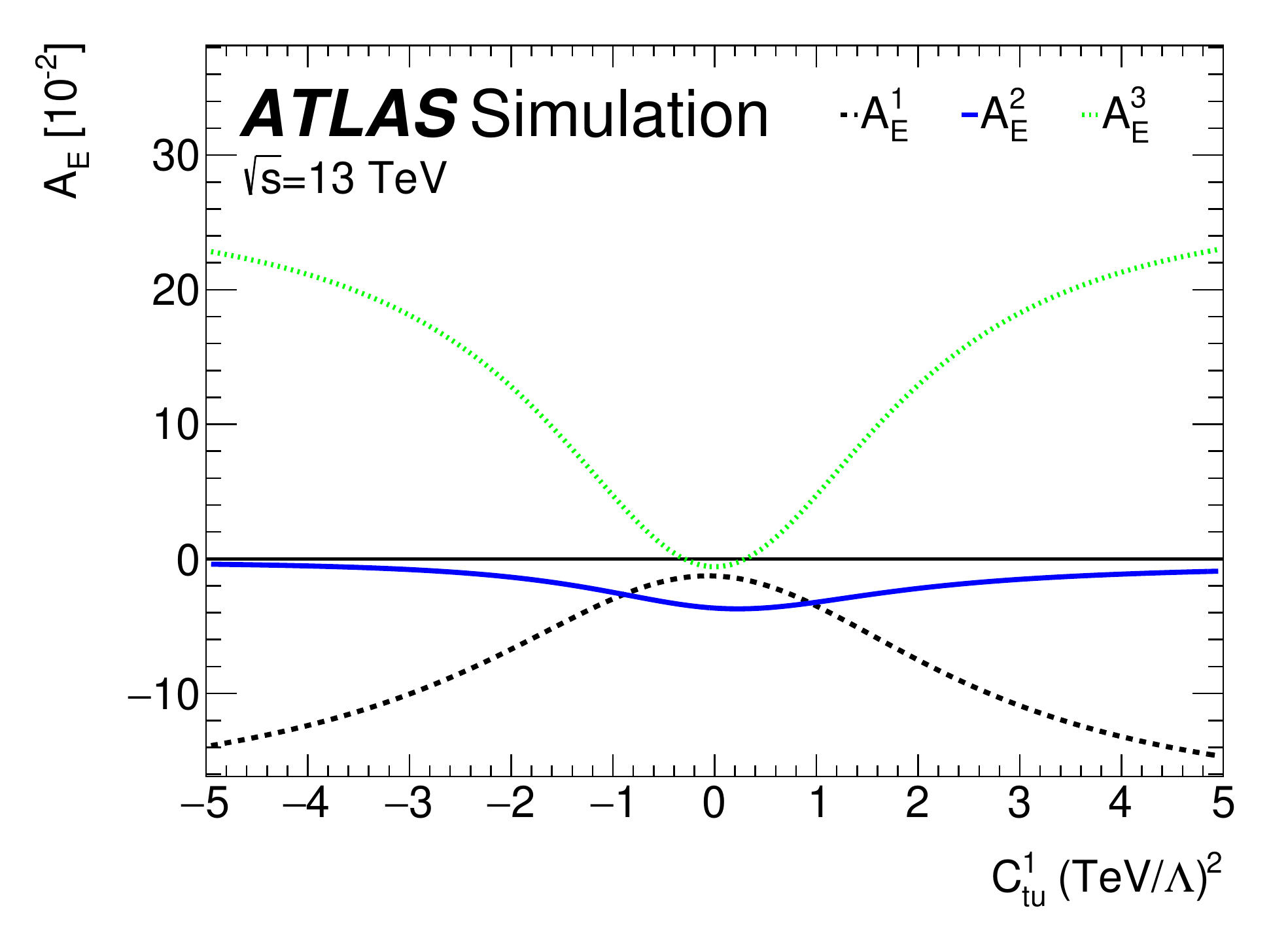}}\hspace*{0.5cm}
\subfloat[]{ \includegraphics[width=.48\textwidth]{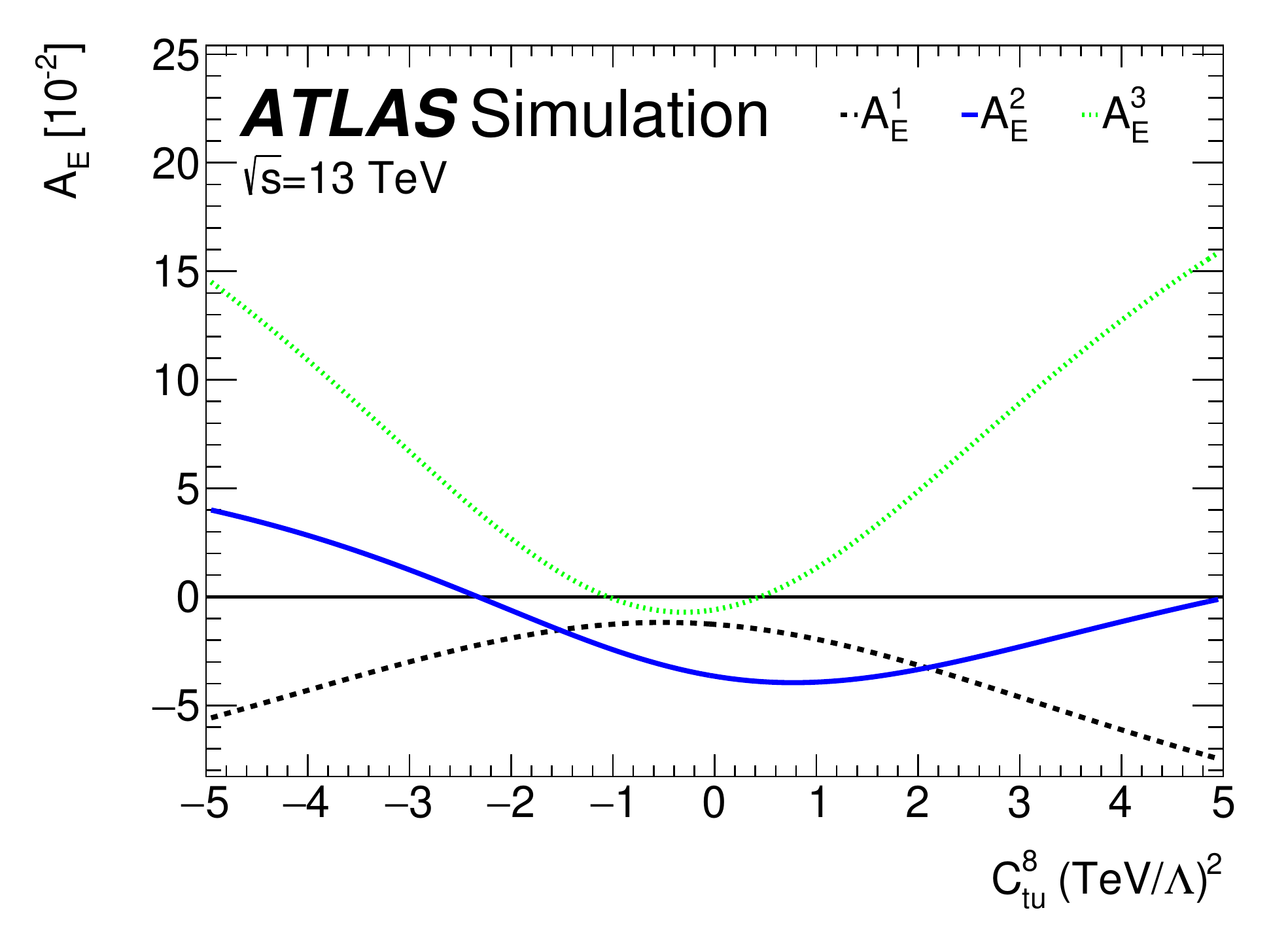}}\\
\caption{Energy asymmetry $A_E$ in three $\theta_j$ bins, $A_E^1$ (black, dashed), $A_E^2$ (blue, solid) and $A_E^3$ (green, dotted), as a function of individual Wilson coefficients. In each sub-figure, only one coefficient is considered, while all other coefficients are set to zero.}
\label{fig:smeft:avc}
\end{figure}

Figure~\ref{fig:smeft:avc} shows the energy asymmetry as a function of individual Wilson coefficients; $A_E^1$, $A_E^2$ and $A_E^3$ correspond to the three bins in $\theta_j$, $[0,\pi/4,3/5\pi,\pi]$, respectively.
To assess the sensitivity of the energy asymmetry to the various operator contributions, $\chi^2$ fits of the SMEFT predictions $x^{\text{SMEFT}}$ to the measured energy asymmetry $x^{\text{d}}$ in the three jet-angle bins $x = (A_E^1,A_E^2,A_E^3)^\top$ are performed, fitting one or two coefficients at a time, using
\begin{equation*}
\chi^2 = (x^{\text{d}} - x^{\text{SMEFT}})^\top C^{-1} (x^{\text{d}} - x^{\text{SMEFT}})\,.
\end{equation*}
Correlations among the measurements and among the predictions are taken into account by the covariance matrix
\begin{equation*}
C = C^{\text{d}} + C^{t}, \qquad C^{t} = C^{\text{MC stat}} + C^{\text{scale}},
\end{equation*}
which contains the measurement uncertainties $C^\text{d}$ and the prediction's theoretical uncertainties $C^\text{t}$ due to both the MC sample size, $C^{\text{MC stat}}$, and the renormalisation and factorisation scale uncertainties, $C^{\text{scale}}$. The predictions $x^{\text{SMEFT}}$ are functions of the Wilson coefficients $C_k$, shown in Eq.~\eqref{eq:obs-smeft}, that enter the fit as parameters. The corresponding uncertainties associated with $\sigma_\text{S,A}^k$ and $\sigma_\text{S,A}^{kl}$ are contained in $C^{\text{MC stat}}$ and $C^{\text{scale}}$. The covariance matrix for the scale uncertainties in the SM and SMEFT contributions to $x^{\text{SMEFT}}$ in each $\theta_j$ bin is obtained from the envelope of nine different settings for the renormalisation and factorisation scales, varied independently by factors of 0.5 and 2.0 and assuming no correlation between the bins. The choice of no bin-to-bin correlation has negligible impact on the result, since theoretical uncertainties have only a small impact on this observable, except in the last $\theta_j$ bin.
 
\begin{figure}[t!]
\centering
\includegraphics[width=.48\textwidth]{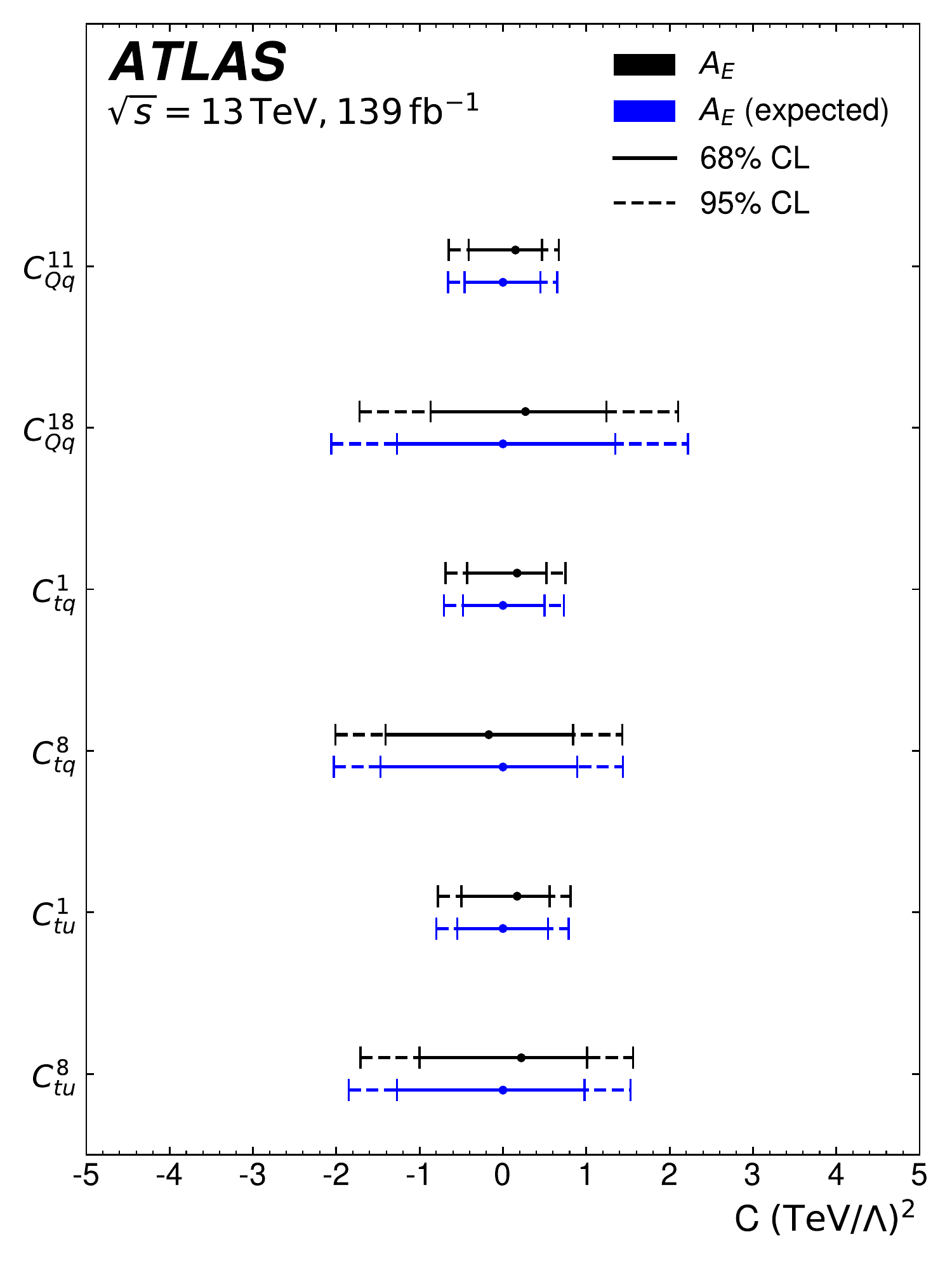}
\caption{Bounds on individual Wilson coefficients from the energy asymmetry, obtained from a combined fit to the measured (black) and expected (blue) energy asymmetry in all three $\theta_j$ bins. The bounds on $C\,(\tev/\Lambda)^2$ are reported at the 68\% CL (solid) and 95\% CL (dashed).
\label{fig:smeft:Limits1D}}
\end{figure}

Figure~\ref{fig:smeft:Limits1D} shows the bounds on individual Wilson coefficients obtained from combined fits to $A_E$ in all three $\theta_j$ bins with only one operator contributing at a time.
The corresponding 68\% and 95\% confidence level (CL) intervals are listed in Table~\ref{tab:smeft:Limits1D}. Overall, the energy asymmetry is sensitive to four-quark operator coefficients in the range of $[-2,2]\,(\text{TeV}/\Lambda)^2$ at 95\% CL, with the precise bounds varying for individual operators.
 
\begin{table}[ht]
\caption{Bounds on individual Wilson coefficients $C\,(\tev/\Lambda)^2$ from the energy asymmetry, obtained from a combined fit to its values in all three $\theta_j$ bins: $A_E^1$, $A_E^2$, and $A_E^3$. The SMEFT predictions  include  operator contributions to $\sigma_\text{S}$ and $\sigma_\text{A}$ up to $\mathcal{O}(\Lambda^{-4})$ (left) and up to $\mathcal{O}(\Lambda^{-2})$ (right).
\vspace{10pt}
\label{tab:smeft:Limits1D}}
\centering
\resizebox{0.8\textwidth}{!}{
\begin{tabular}{c|cc|cc}
\multirow{2}{*}{$C\,(\tev/\Lambda)^2$} & \multicolumn{2}{c}{$\ea$ ($\Lambda^{-4}$)} & \multicolumn{2}{c}{$\ea$  ($\Lambda^{-2}$)} \\
& 68\% CL & 95\% CL & 68\% CL & 95\% CL \\
\toprule
$C_{Qq}^{11}$ & $[-0.41,0.47]$ & $[-0.65,0.67]$ & $[-0.68,4.06]$ & $[-3.36,6.16]$\\
$C_{Qq}^{18}$ & $[-0.87,1.24]$ & $[-1.72,2.10]$ & $[-1.26,4.76]$ & $[-3.24,9.64]$\\
$C_{tq}^{1}$ & $[-0.43,0.52]$ & $[-0.69,0.75]$ & $[-0.60,5.76]$ & $[-3.42,9.36]$\\
$C_{tq}^{8}$ & $[-1.41,0.84]$ & $[-2.01,1.43]$ & $[-1.86,1.70]$ & $[-3.30,3.98]$\\
$C_{tu}^{1}$ & $[-0.50,0.56]$ & $[-0.78,0.81]$ & $[-0.96,5.82]$ & $[-4.72,8.88]$\\
$C_{tu}^{8}$ & $[-1.00,1.01]$ & $[-1.71,1.56]$ & $[-1.30,2.52]$ & $[-3.02,4.66]$\\
\bottomrule
\end{tabular}
} 
\end{table}
 
To estimate the relative impact of $\mathcal{O}(\Lambda^{-2})$ versus $\mathcal{O}(\Lambda^{-4})$ operator contributions, Table~\ref{tab:smeft:Limits1D} shows the bounds obtained from the energy asymmetry when including operator contributions to the charge-symmetric and charge-asymmetric cross sections $\sigma_\text{S}$ and $\sigma_\text{A}$ up to $\mathcal{O}(\Lambda^{-4})$ (left two columns) and up to $\mathcal{O}(\Lambda^{-2})$ (right two columns). Comparing these two scenarios, it is clear that the bounds are dominated by the $\mathcal{O}(\Lambda^{-4})$ contributions $\sigma_\text{S,A}^{kl}$. The results at $\mathcal{O}(\Lambda^{-2})$ are listed here only to assess the relevance of the various operator contributions.
 
Due to its sensitivity to four-quark operators the energy asymmetry will provide additional constraining power when included in global SMEFT fits. Recent global fits of top-quark observables lead to stronger bounds on four-quark coefficients, see for instance Figure~5.7 in \Bibref{Hartland:2019bjb}, Figure~13 in \Bibref{Brivio:2019ius} or Figure~5 in \Bibref{Ellis:2020unq}. This is due to charge-symmetric observables such as the invariant mass distribution $\text{d}\sigma/\text{d} m_{t\bar t}$, where the sensitivity to four-quark operators is strongly enhanced at high energy.
 
\begin{figure}
\subfloat[]{\includegraphics[width=.45\textwidth]{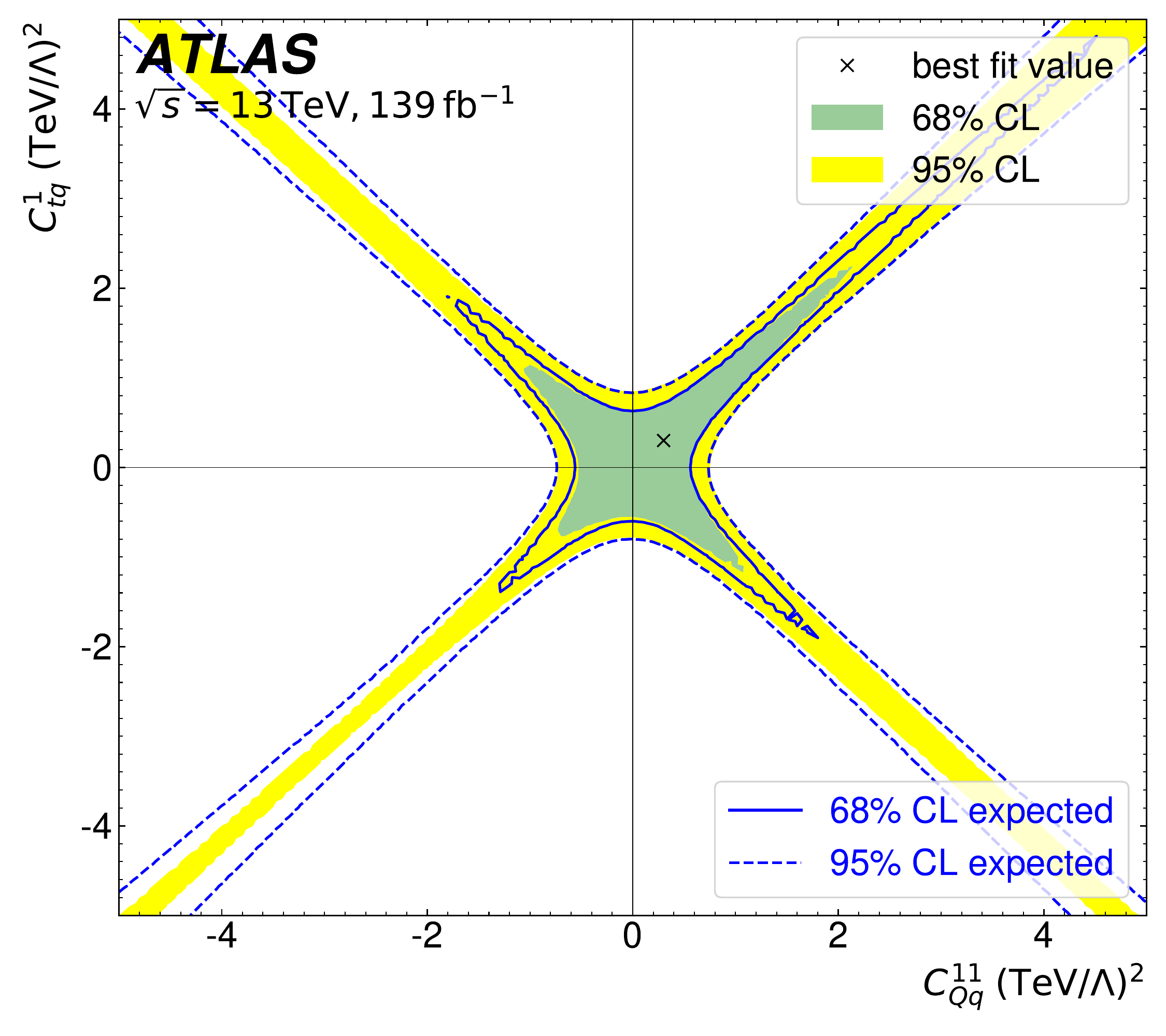}\label{fig:smeft:Limits2D:Ae_diff_Ae_diff_asimov:a}}\hspace*{0.5cm}
\subfloat[]{\includegraphics[width=.45\textwidth]{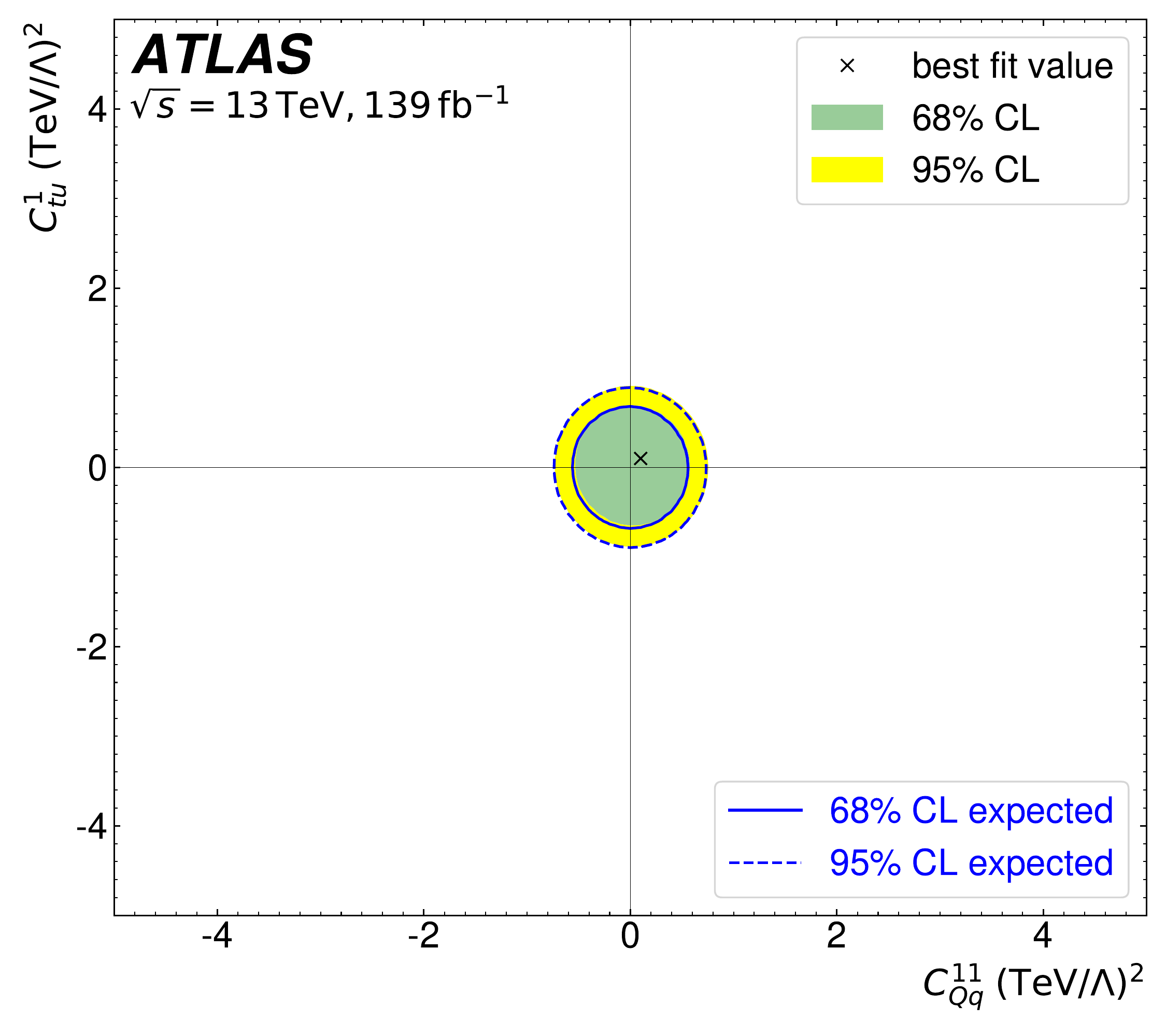}\label{fig:smeft:Limits2D:Ae_diff_Ae_diff_asimov:b}}\\
\vspace{-0.25cm}
\subfloat[]{\includegraphics[width=.45\textwidth]{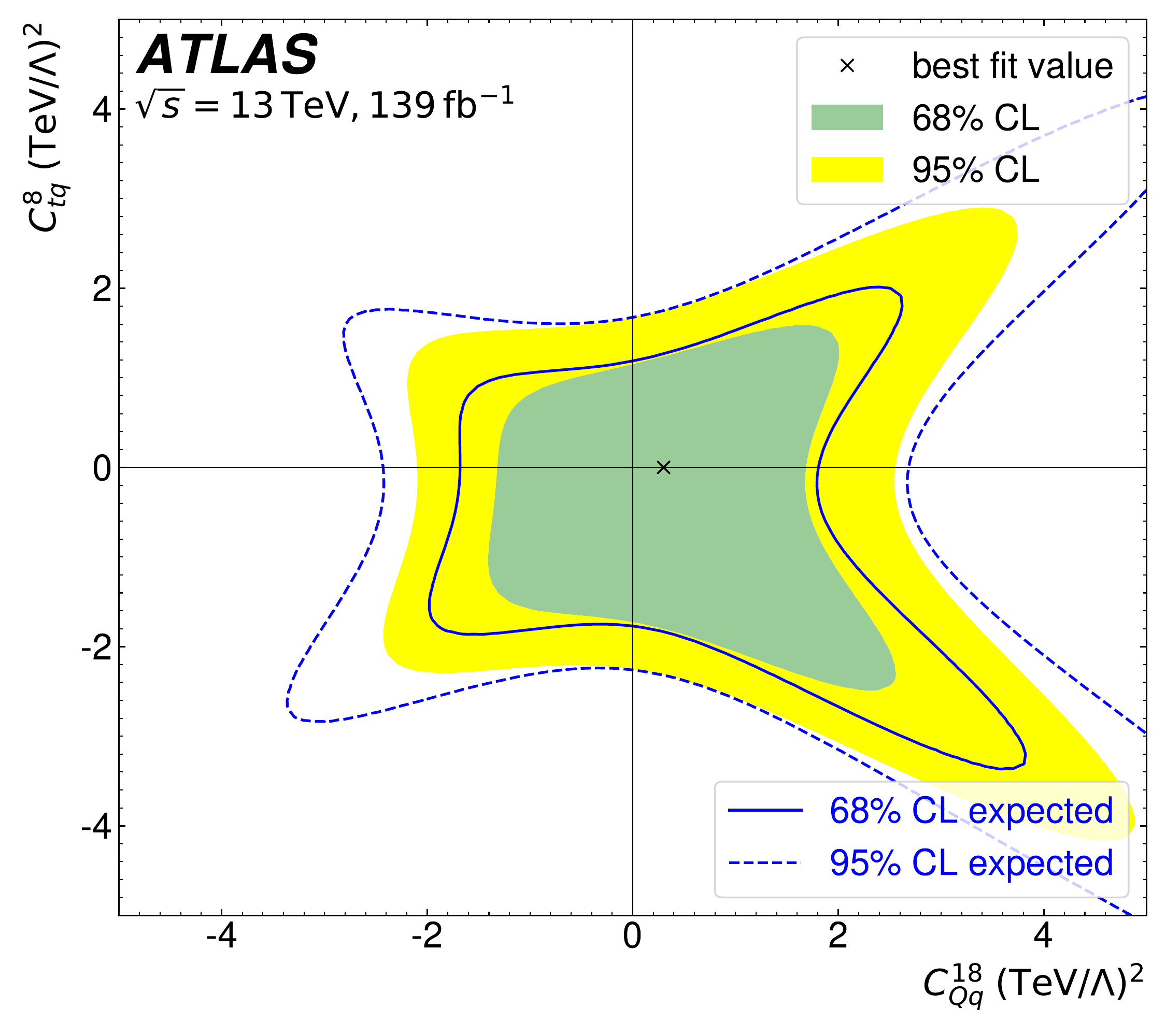}\label{fig:smeft:Limits2D:Ae_diff_Ae_diff_asimov:c}}\hspace*{0.5cm}
\subfloat[]{\includegraphics[width=.45\textwidth]{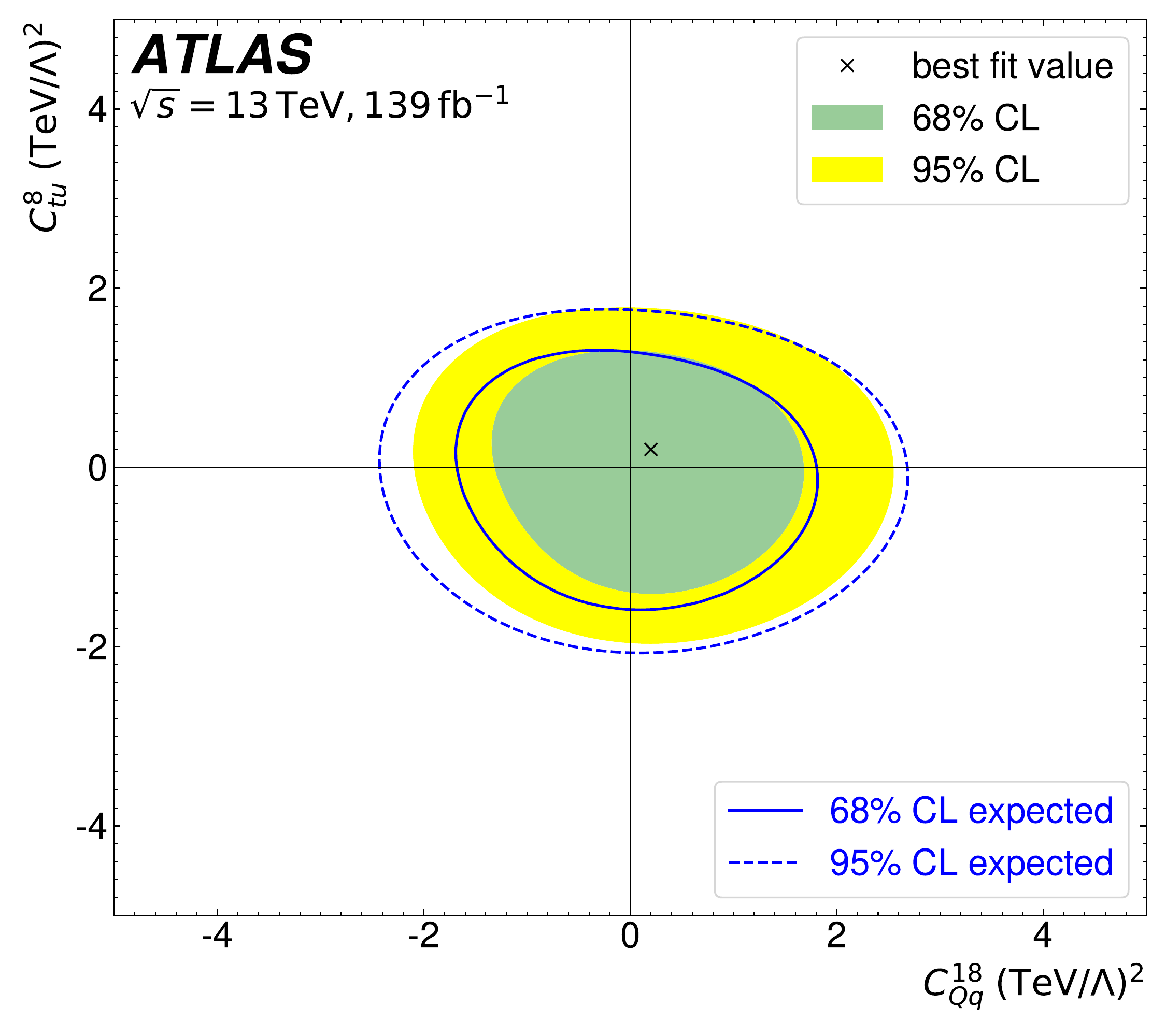}\label{fig:smeft:Limits2D:Ae_diff_Ae_diff_asimov:d}}\\
\vspace{-0.25cm}
\subfloat[]{\includegraphics[width=.45\textwidth]{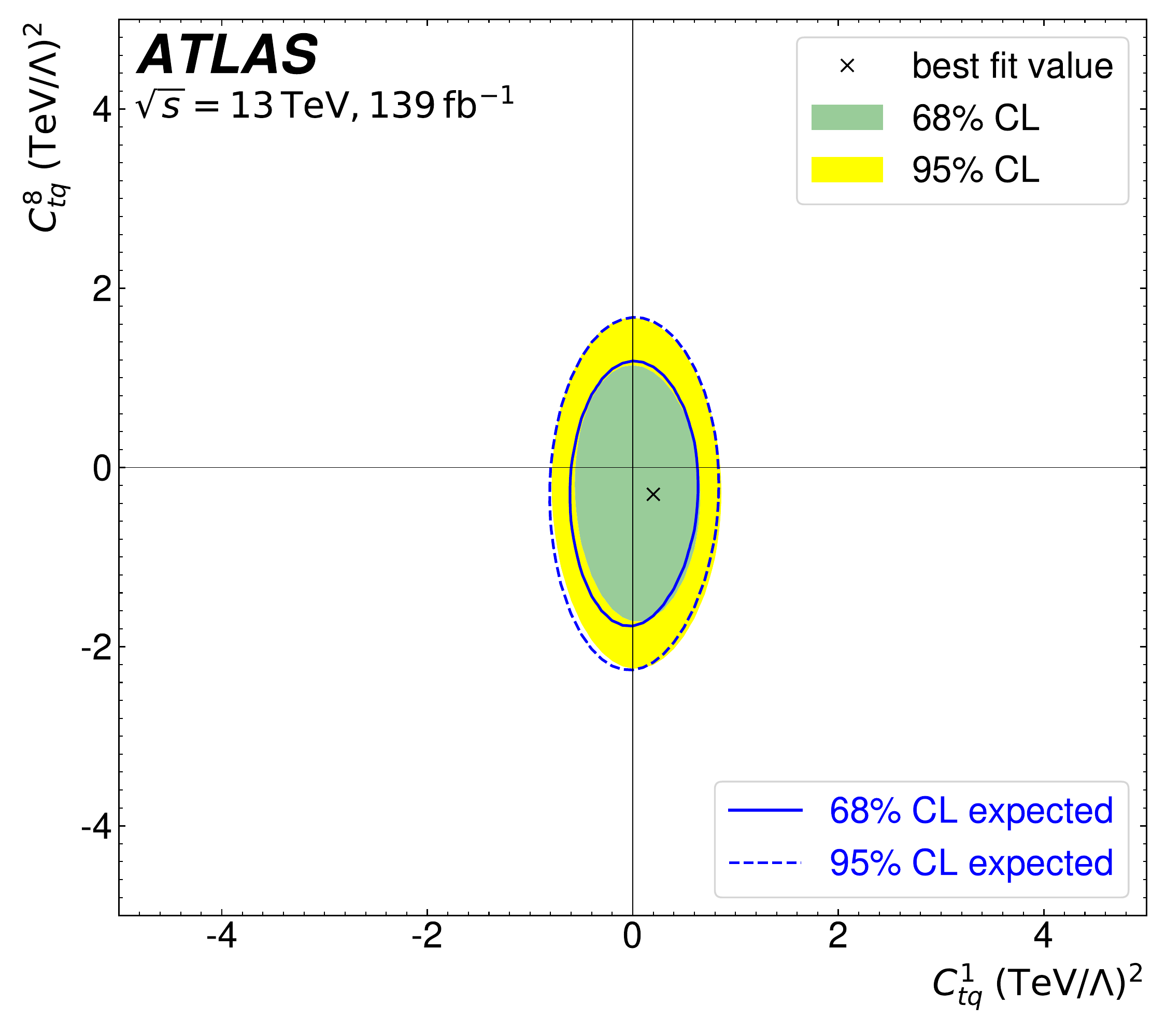}\label{fig:smeft:Limits2D:Ae_diff_Ae_diff_asimov:e}}\hspace*{0.5cm}
\subfloat[]{\includegraphics[width=.45\textwidth]{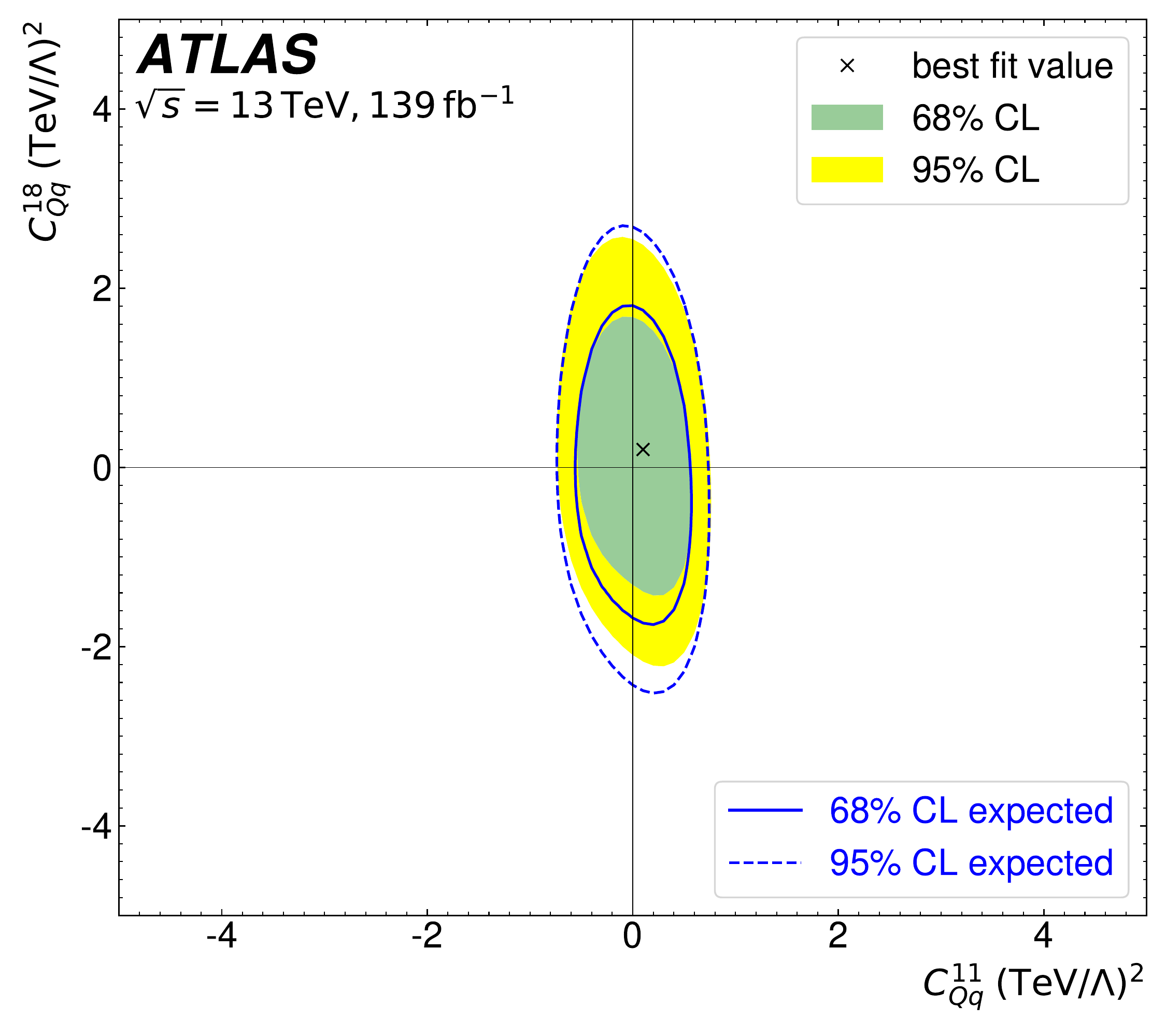}\label{fig:smeft:Limits2D:Ae_diff_Ae_diff_asimov:f}}\\
\caption{Bounds on Wilson coefficients from two-parameter fits of the energy asymmetry \ea in all three $\theta_j$ bins. The inner green and outer yellow areas show the 68\% and 95\% CL limits obtained from the measured energy asymmetry, respectively. The black cross denotes the minimum of the $\chi^2$ fit. Solid and dashed blue contours show the expected 68\% and 95\% CL limits from the energy asymmetry, respectively.\label{fig:smeft:Limits2D:Ae_diff_Ae_diff_asimov}}
\end{figure}

The main asset of the energy asymmetry is its ability to resolve new directions in the SMEFT parameter space. In particular, charge asymmetries are very sensitive to the quark chiralities and the colour structure of effective interactions. The analytic dependence of the energy asymmetry has been studied in detail in \Bibref{Basan:2020btr}. To explore the numerical sensitivity to the chiral and colour structure of the six operators from Eq.~\eqref{eq:operators}, fits to pairs of Wilson coefficients were performed. Figure~\ref{fig:smeft:Limits2D:Ae_diff_Ae_diff_asimov} shows the bounds on these coefficients from a combined fit of the energy asymmetry in all three jet-angle bins. Most bounds are centred near $(0,0)$, due to good agreement between the measurement and the SM prediction.
 
The various pairs of operators have been chosen in order to investigate their chiral structure and colour structure separately. Figures~\ref{fig:smeft:Limits2D:Ae_diff_Ae_diff_asimov:a} and \ref{fig:smeft:Limits2D:Ae_diff_Ae_diff_asimov:b} show colour-singlet operators with different quark chiralities. Figures~\ref{fig:smeft:Limits2D:Ae_diff_Ae_diff_asimov:c} and~\ref{fig:smeft:Limits2D:Ae_diff_Ae_diff_asimov:d} show the same chirality scenarios as in Figures~\ref{fig:smeft:Limits2D:Ae_diff_Ae_diff_asimov:a} and~\ref{fig:smeft:Limits2D:Ae_diff_Ae_diff_asimov:b}, but for colour-octet operators. Figures~\ref{fig:smeft:Limits2D:Ae_diff_Ae_diff_asimov:e} and \ref{fig:smeft:Limits2D:Ae_diff_Ae_diff_asimov:f} show operators that differ only in their colour structure. The main features observed are discussed below; details about the operator dependence of the asymmetry can be found in \Bibref{Basan:2020btr}.
 
In Figures \ref{fig:smeft:Limits2D:Ae_diff_Ae_diff_asimov:a} and \ref{fig:smeft:Limits2D:Ae_diff_Ae_diff_asimov:b}, the two operators differ only by the chirality of the top quark (left) and the chirality of the up quark (right). For operators with left- versus right-handed tops (left), a blind direction is observed along $C_{Qq}^{11} = \pm C_{tq}^1$. This blind direction is slightly broken by the linear SMEFT contribution of $\mathcal{O}(\Lambda^{-2})$ (see Equation~(3.17) in \Bibref{Basan:2020btr}). For operators with different initial quark chiralities (right), the blind direction is broken at $\mathcal{O}(\Lambda^{-4})$, because the operators $O_{Qq}^{1,1}$ and $O_{tu}^{1}$ do not interfere with each other in the limit of massless incoming quarks. Colour-octet operators, shown in Figures  \ref{fig:smeft:Limits2D:Ae_diff_Ae_diff_asimov:c} and  \ref{fig:smeft:Limits2D:Ae_diff_Ae_diff_asimov:d}, interfere with QCD. This effect lifts the blind direction in the $(C_{Qq}^{18},C_{tq}^8)$ plane.
 
The bounds for colour singlets are generally stronger than for colour octets. This can be seen by comparing the bounds in Figures~\ref{fig:smeft:Limits2D:Ae_diff_Ae_diff_asimov:a} and~\ref{fig:smeft:Limits2D:Ae_diff_Ae_diff_asimov:b} with those in Figures~\ref{fig:smeft:Limits2D:Ae_diff_Ae_diff_asimov:c} and~\ref{fig:smeft:Limits2D:Ae_diff_Ae_diff_asimov:d} or by comparing the bounds on each of the coefficients in Figures~\ref{fig:smeft:Limits2D:Ae_diff_Ae_diff_asimov:e} and~\ref{fig:smeft:Limits2D:Ae_diff_Ae_diff_asimov:f}. The different sensitivity is due to colour factors in the amplitude, which lead to a larger contribution of colour-singlet coefficients to the asymmetry.
 
Due to the extra jet in $t\bar t j$ production, the QCD structure of the energy asymmetry is not the same as for the rapidity asymmetry in $t\bar t$ production. Indeed, the two asymmetries probe different directions in chiral and colour space. This complementarity can be estimated by comparing Figure~\ref{fig:smeft:Limits2D:Ae_diff_Ae_diff_asimov} with Figure~6 in \Bibref{Basan:2020btr}, which shows the expected bounds from the energy asymmetry compared with the expected bounds from the rapidity asymmetry. For the colour-singlet operators in Figures~\ref{fig:smeft:Limits2D:Ae_diff_Ae_diff_asimov:a} and~\ref{fig:smeft:Limits2D:Ae_diff_Ae_diff_asimov:b}, the two asymmetries probe similar areas in the parameter space. For the colour-octet operators in Figures~\ref{fig:smeft:Limits2D:Ae_diff_Ae_diff_asimov:c} and~\ref{fig:smeft:Limits2D:Ae_diff_Ae_diff_asimov:d}, however, the shapes of the bounds look very different. In particular, the rapidity asymmetry leaves a blind direction in the $(C_{Qq}^{18},C_{tq}^8)$ plane, which is broken by the energy asymmetry due to operator interference with the QCD amplitude as shown in Figure~\ref{fig:smeft:Limits2D:Ae_diff_Ae_diff_asimov:c}. The different shapes of the bounds in Figures~\ref{fig:smeft:Limits2D:Ae_diff_Ae_diff_asimov:e} and~\ref{fig:smeft:Limits2D:Ae_diff_Ae_diff_asimov:f} are due to the colour-singlet and colour-octet contributions differing between $t\bar{t}$ and $ttj$ production, which is probed by the asymmetries with a high sensitivity.
 
This comparison demonstrates once more the relevance of the energy asymmetry for global fits. Combining the energy and rapidity asymmetries in a joint fit would result in a better resolution of the Wilson coefficients in the SMEFT parameter space. Moreover, the energy asymmetry resolves remaining blind or nearly blind directions left by the rapidity asymmetry and other top-quark observables. In global analyses the energy asymmetry can thus be a useful tool to gain information about the chiral and colour structure of possible UV theories.
 
 
\FloatBarrier
 
\section{Summary}
\label{sec:summary}
This paper reports the first measurement of the energy asymmetry in jet-associated top-quark pair production in the boosted topology in the lepton+jets decay channel, using $139\ifb$ of $\sqrt{s}=13\tev$ proton-proton collision data collected by the ATLAS experiment at the LHC. The measurement was performed differentially in three bins of the jet scattering angle $\theta_j$. A fully Bayesian unfolding method was used to correct for detector resolution and acceptance effects.
 
The differential measurements of the energy asymmetry in all three $\theta_j$ bins are consistent with the SM predictions calculated at NLO in QCD, including effects of parton showering and hadronisation. In the second bin, $\pi/4 < \theta_j < 3\pi/5$, with jet emission perpendicular to the beam axis, the energy asymmetry is measured as $A_E^2 = -0.043\pm0.020$, in good agreement with the QCD prediction of $-0.037\pm 0.003$. The measurement uncertainty is dominated by the limited data sample in the boosted phase-space region.
 
To determine the sensitivity of the energy asymmetry to new physics, this observable was analyzed within the SMEFT framework. The energy asymmetry is highly sensitive to the chiral and colour structure of four-quark operators with top quarks. Based on SMEFT predictions in terms of dimension-six operators, bounds on the Wilson coefficients of individual operators are extracted.
 
The energy asymmetry in $\ttj$ production has a sensitivity to some individual four-quark operators that is similar to that of other observables in the top sector and probes new directions in the parameter space of Wilson coefficients. The sensitivity to these new directions is illustrated in the two-parameter fits of operator pairs with different chiral and colour structures. These results show that the energy asymmetry complements the sensitivity of other observables, such as differential cross sections and the rapidity asymmetry, and hence that the energy asymmetry is a valuable new observable in global SMEFT fits.


\section*{Acknowledgements}

We thank CERN for the very successful operation of the LHC, as well as the
support staff from our institutions without whom ATLAS could not be
operated efficiently.
 
We acknowledge the support of
ANPCyT, Argentina;
YerPhI, Armenia;
ARC, Australia;
BMWFW and FWF, Austria;
ANAS, Azerbaijan;
SSTC, Belarus;
CNPq and FAPESP, Brazil;
NSERC, NRC and CFI, Canada;
CERN;
ANID, Chile;
CAS, MOST and NSFC, China;
Minciencias, Colombia;
MEYS CR, Czech Republic;
DNRF and DNSRC, Denmark;
IN2P3-CNRS and CEA-DRF/IRFU, France;
SRNSFG, Georgia;
BMBF, HGF and MPG, Germany;
GSRI, Greece;
RGC and Hong Kong SAR, China;
ISF and Benoziyo Center, Israel;
INFN, Italy;
MEXT and JSPS, Japan;
CNRST, Morocco;
NWO, Netherlands;
RCN, Norway;
MEiN, Poland;
FCT, Portugal;
MNE/IFA, Romania;
JINR;
MES of Russia and NRC KI, Russian Federation;
MESTD, Serbia;
MSSR, Slovakia;
ARRS and MIZ\v{S}, Slovenia;
DSI/NRF, South Africa;
MICINN, Spain;
SRC and Wallenberg Foundation, Sweden;
SERI, SNSF and Cantons of Bern and Geneva, Switzerland;
MOST, Taiwan;
TAEK, Turkey;
STFC, United Kingdom;
DOE and NSF, United States of America.
In addition, individual groups and members have received support from
BCKDF, CANARIE, Compute Canada and CRC, Canada;
COST, ERC, ERDF, Horizon 2020 and Marie Sk{\l}odowska-Curie Actions, European Union;
Investissements d'Avenir Labex, Investissements d'Avenir Idex and ANR, France;
DFG and AvH Foundation, Germany;
Herakleitos, Thales and Aristeia programmes co-financed by EU-ESF and the Greek NSRF, Greece;
BSF-NSF and GIF, Israel;
Norwegian Financial Mechanism 2014-2021, Norway;
NCN and NAWA, Poland;
La Caixa Banking Foundation, CERCA Programme Generalitat de Catalunya and PROMETEO and GenT Programmes Generalitat Valenciana, Spain;
G\"{o}ran Gustafssons Stiftelse, Sweden;
The Royal Society and Leverhulme Trust, United Kingdom.
 
The crucial computing support from all WLCG partners is acknowledged gratefully, in particular from CERN, the ATLAS Tier-1 facilities at TRIUMF (Canada), NDGF (Denmark, Norway, Sweden), CC-IN2P3 (France), KIT/GridKA (Germany), INFN-CNAF (Italy), NL-T1 (Netherlands), PIC (Spain), ASGC (Taiwan), RAL (UK) and BNL (USA), the Tier-2 facilities worldwide and large non-WLCG resource providers. Major contributors of computing resources are listed in Ref.~\cite{ATL-SOFT-PUB-2021-003}.
 

\printbibliography
 
\clearpage
 
\begin{flushleft}
\hypersetup{urlcolor=black}
{\Large The ATLAS Collaboration}

\bigskip

\AtlasOrcid[0000-0002-6665-4934]{G.~Aad}$^\textrm{\scriptsize 98}$,    
\AtlasOrcid[0000-0002-5888-2734]{B.~Abbott}$^\textrm{\scriptsize 124}$,    
\AtlasOrcid[0000-0002-7248-3203]{D.C.~Abbott}$^\textrm{\scriptsize 99}$,    
\AtlasOrcid[0000-0002-2788-3822]{A.~Abed~Abud}$^\textrm{\scriptsize 34}$,    
\AtlasOrcid[0000-0002-1002-1652]{K.~Abeling}$^\textrm{\scriptsize 51}$,    
\AtlasOrcid[0000-0002-2987-4006]{D.K.~Abhayasinghe}$^\textrm{\scriptsize 91}$,    
\AtlasOrcid[0000-0002-8496-9294]{S.H.~Abidi}$^\textrm{\scriptsize 27}$,    
\AtlasOrcid[0000-0002-9987-2292]{A.~Aboulhorma}$^\textrm{\scriptsize 33e}$,    
\AtlasOrcid[0000-0001-5329-6640]{H.~Abramowicz}$^\textrm{\scriptsize 157}$,    
\AtlasOrcid[0000-0002-1599-2896]{H.~Abreu}$^\textrm{\scriptsize 156}$
% ...
\bigskip
\\

% ...
$^{27}$Physics Department, Brookhaven National Laboratory, Upton NY; United States of America.\\
% ...
$^{33}$$^{(a)}$Facult\'e des Sciences Ain Chock, R\'eseau Universitaire de Physique des Hautes Energies - Universit\'e Hassan II, Casablanca;$^{(b)}$Facult\'{e} des Sciences, Universit\'{e} Ibn-Tofail, K\'{e}nitra;$^{(c)}$Facult\'e des Sciences Semlalia, Universit\'e Cadi Ayyad, LPHEA-Marrakech;$^{(d)}$LPMR, Facult\'e des Sciences, Universit\'e Mohamed Premier, Oujda;$^{(e)}$Facult\'e des sciences, Universit\'e Mohammed V, Rabat;$^{(f)}$Mohammed VI Polytechnic University, Ben Guerir; Morocco.\\
$^{34}$CERN, Geneva; Switzerland.\\
% ...
$^{51}$II. Physikalisches Institut, Georg-August-Universit\"{a}t G\"ottingen, G\"ottingen; Germany.\\
% ...
$^{91}$Department of Physics, Royal Holloway University of London, Egham; United Kingdom.\\
% ...
$^{98}$CPPM, Aix-Marseille Universit\'e, CNRS/IN2P3, Marseille; France.\\
$^{99}$Department of Physics, University of Massachusetts, Amherst MA; United States of America.\\
% ...
$^{124}$Homer L. Dodge Department of Physics and Astronomy, University of Oklahoma, Norman OK; United States of America.\\
% ...
$^{156}$Department of Physics, Technion, Israel Institute of Technology, Haifa; Israel.\\
$^{157}$Raymond and Beverly Sackler School of Physics and Astronomy, Tel Aviv University, Tel Aviv; Israel.\\
% ...
$^{a}$ Also at Borough of Manhattan Community College, City University of New York, New York NY; United States of America.\\
$^{b}$ Also at Bruno Kessler Foundation, Trento; Italy.\\
$^{c}$ Also at Center for High Energy Physics, Peking University; China.\\
$^{d}$ Also at Centro Studi e Ricerche Enrico Fermi; Italy.\\
$^{e}$ Also at CERN, Geneva; Switzerland.\\
$^{f}$ Also at D\'epartement de Physique Nucl\'eaire et Corpusculaire, Universit\'e de Gen\`eve, Gen\`eve; Switzerland.\\
% ...

\end{flushleft}


\end{document}